\documentclass[twocolumn]{emulateapj}

\def\alwaysmath#1{\ifmmode{#1}\else{$#1$}\fi}

\lefthead{Law et al. 2006}
\righthead{Rest-UV Galaxy Morphology}

\begin{document}
\title{The Physical Nature of Rest-UV Galaxy Morphology During the Peak Epoch of Galaxy Formation}
\author{\sc David R. Law\altaffilmark{1}, Charles C. Steidel\altaffilmark{1}, Dawn K. Erb\altaffilmark{2}, Max Pettini\altaffilmark{3},
Naveen A. Reddy\altaffilmark{1}, Alice E. Shapley\altaffilmark{4}, Kurt L. Adelberger\altaffilmark{5}, and David J. Simenc\altaffilmark{1}}
\altaffiltext{1}{Department of Astronomy, California Institute of Technology, MS 105-24,
Pasadena, CA 91125 (drlaw, ccs, nar@astro.caltech.edu, dsimenc@caltech.edu)}
\altaffiltext{2}{Harvard-Smithsonian Center for Astrophysics, MS 20, 60 Garden St, Cambridge, MA 02138 (derb@cfa.harvard.edu)}
\altaffiltext{3}{Institute of Astronomy, Madingley Road, Cambridge CB3 0HA, UK (pettini@ast.cam.ac.uk)}
\altaffiltext{4}{Department of Astrophysical Sciences, Princeton University, Peyton Hall, Ivy Lane, Princeton, NJ 08544 (aes@astro.princeton.edu)}
\altaffiltext{5}{McKinsey and Company, 1420 Fifth Avenue, Suite 3100, Seattle, WA 98101 (kurt@ociw.edu)}

\begin{abstract}
Motivated by the irregular and little-understood morphologies of $z \sim 2 - 3$ galaxies, we use non-parametric
coefficents to quantify the morphologies of 216 galaxies which have been spectroscopically
confirmed to lie at redshifts $z =$ 1.8 - 3.4 in the GOODS-N field.  Using measurements of
ultraviolet (UV) and optical spectral lines, multi-band photometric data, and stellar population models we statistically
assess possible correlations between galaxy morphology and physical observables such as stellar mass,
star formation rate, and the strength of galaxy-scale outflows.
We find evidence that dustier galaxies have more nebulous UV morphologies
and that larger, more luminous galaxies may drive stronger outflows, but otherwise conclude that UV
morphology is either statistically decoupled from the majority of physical observables or determined by too
complex a combination of physical processes to provide characterizations with predictive power.
Given the absence of strong correlations between UV morphology and physical parameters such as star formation rates,
we are therefore unable to support the hypothesis that morphologically irregular galaxies predominantly represent
major galaxy mergers.
Comparing galaxy samples, we find that IR-selected $BzK$ galaxies and radio-selected submillimeter galaxies (SMGs) have 
UV morphologies similar to the optically selected sample, while distant red galaxies (DRGs) are more nebulous.

\end{abstract}

\keywords{cosmology: observations ---  galaxies: fundamental parameters --- galaxies: high-redshift --- galaxies: irregular --- 
galaxies: starburst --- galaxies: structure}

\section{INTRODUCTION}

In the local universe the projected distribution of luminous matter within a galaxy, i.e. the morphology,
often provides a wealth of information about that galaxy's kinematics, rate of star formation,
and recent merger history.  In the classical picture, late-type spiral galaxies harbor active star formation in
the gas-rich arms of a flattened rotating disk, while early-type elliptical galaxies
tend to be more massive, dispersion supported, and quiescent systems.
At high redshifts from $z \sim 2 - 3$ however the morphologies of typical
galaxies are highly irregular (Abraham et al. 1996, Kajisawa \& Yamada 2001, Conselice et al. 2005), 
frequently composed of multiple spatially separated components, and appear to bear little similarity to the local 
Hubble-type population.
It is uncertain whether these irregular morphologies are due to patchy star formation, prevalent merger activity,
or some other physical process and consequently unknown whether these morphologies can 
(analogously to local galaxies) tell
us anything about the star formation rate, mass, or stellar kinematics of galaxies at high redshifts.

Since morphological studies are often performed at optical wavelengths which probe rest-frame ultraviolet (UV) radiation 
for galaxies at redshifts $z \gtrsim 1$,
one might {\it expect} that the morphologies of such galaxies should appear irregular since radiation at such wavelengths
predominantly traces emission from the brightest active star forming regions rather than the redder bulk of the
stellar population (Dickinson 2000).  UV emission tends to be patchy and irregular even for local Hubble-type
galaxies (e.g. Gordon et al. 2004), as
in the case of the local galaxy merger VV~114 (whose broad rest-UV absorption line spectra suggest that
it may be a local analog to $z \sim 2 - 3$ Lyman Break Galaxies; Grimes et al. 2006) whose near-infrared (NIR) morphology clearly
shows a pair of interacting late-type galaxies while the rest-UV morphology shows only scattered clumps of emission (Goldader et al. 2002).
However, high-redshift galaxies have irregular morphologies not only in the rest-UV, but often at rest-optical wavelengths as well
(Dickinson 2000; Papovich et al. 2005), indicating that (in contrast to local galaxies) 
both wavelength regimes are dominated by emission from young starbursting components
and therefore that there may be some fundamental difference between the two samples.

One popular explanation for these multi-component, irregular morphologies is that they represent major merger systems, 
and that their prevalence indicates 
that the rate of major mergers was much greater at high redshifts than in the local universe (e.g. Conselice et al. 2003).
Such a conclusion fits well within the framework of cold dark matter (CDM) theory, and may additionally be supported by stellar population
analyses (e.g. Dickinson et al. 2003) which suggest that many galaxies in the local universe accumulated a large fraction of their stellar mass 
at $z \sim 2 - 3$
as might be expected if star formation peaked in this epoch as a result of tidally induced collapse spurred by major mergers.
However, the interpretation of a multi-component or otherwise irregular morphology is not always clear.  In the case of VV~114,
near-IR imaging (Goldader et al. 2002) indicates that all of the clumps of UV emission are associated with only one galaxy of the merger pair
and that the multi-component UV morphology therefore directly reflects clumpy star formation rather than tracing tidally distorted features
from each of the two galaxies.

Building on a body of literature characterizing the morphologies of galaxies at redshifts $z \sim 2 - 3$ (e.g. Abraham et al. 2003; Conselice et al. 2003;
Lotz et al. 2004, 2006; Ravindranath et al. 2006) it is worthwhile to ask whether rest-UV morphologies correspond to any other physical observables
such as UV/optical spectral line strengths 
(e.g. Shapley et al. 2003; Erb et al. 2006a), stellar population models (e.g. Shapley et al. 2005; Erb et al. 2006a; Reddy et al. 2006a),
or rest-optical to IR properties (e.g. Reddy et al. 2006a, 2006b).
In this work, we use non-parametric coefficients to characterize the morphologies of
216 spectroscopically confirmed galaxies in the redshift range $z = 1.8 - 3.4$,
assess the statistical significance of correlations with spectrophotometric 
observables, and discuss the resulting physical interpretation of galaxy morphology.
In \S 2 we describe our galaxy sample and give a basic description of the sample population.
In \S 3 we outline our morphological parameters, comparing our results to the recent studies of Conselice et al. (2003) and Lotz et al. (2004, 2006).
Rest-frame UV spectra are introduced in \S 4, correlations between morphology and spectral line strength
and kinematics are discussed in \S 5.  In \S 6 and \S 7 we compare morphologies with stellar population models
derived from UV to mid-IR photometric data, as well as discussing differences between different samples of high-redshift galaxies and AGN
selected on the basis of various photometric criteria.
Finally, we discuss the implications of our results for the physical interpretation of galaxy morphologies in \S 8.
Our morphological statistics and ancillary data are made publically available in an electronic database located at
http://www.astro.caltech.edu/$\sim$drlaw/GOODS/\\

We assume a standard $\Lambda$CDM cosmology in which $H_0 = 71$ km s$^{-1}$ Mpc$^{-1}$, $\Omega_{\rm m} = 0.27$, and $\Omega_{\rm \Lambda} = 0.73$.

\section{SAMPLE SELECTION}
Our sample is drawn from rest-UV color-selected catalogs of $z \sim 2-3$ star-forming galaxy candidates (Steidel et al. 2003, 2004; 
Adelberger et al. 2004) in the GOODS-N field.  These catalogs are based on deep ground-based imaging, and therefore
select galaxies independent of morphology or surface brightness since even the largest galaxies are nearly unresolved in these
seeing-limited images.  We consider only those galaxies which have been spectroscopically
confirmed to lie in the redshift intervals $z = 1.8 - 2.6$ or $z = 2.6 - 3.4$ (i.e. the peak redshift ranges defined by the selection functions
of the color selection criteria, see Adelberger et al. 2004) and
which exhibit no obvious spectroscopic signatures of active galactic nuclei.
  
\begin{figure}
\plotone{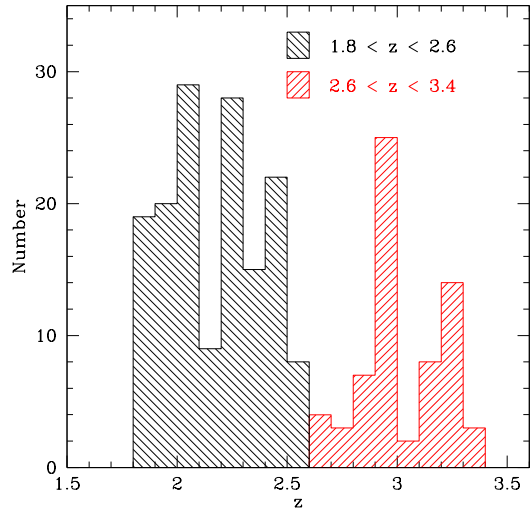}
\caption{Distribution of galaxies with spectroscopic redshift $z$ in the $z \sim 2$ and $z \sim 3$ galaxy samples.}
\label{zdist.fig}
\end{figure}

The redshift distribution of galaxies in our sample is shown in Figure \ref{zdist.fig}: the $z \sim 2$ sample contains 150 galaxies 
in the range $1.8 < z < 2.6$ with mean $\bar{z} = 2.17 \pm 0.21$, while the $z \sim 3$ sample contains 66 galaxies
in the range $2.6 < z < 3.4$ with $\bar{z} = 3.02 \pm 0.19$.

\section{MORPHOLOGIES}

Morphological parameters were determined from deep {\it HST}-ACS imaging obtained as part of the GOODS-N survey (Giavalisco et al. 2004) in
F435W ($B$), F606W ($V$), F775W ($I$), and F850LP ($z$) bandpasses with drizzled 
pixel scale of 50 mas pixel$^{-1}$ and 10$\sigma$ limiting point source
sensitivities of 27.8, 27.8, 27.1, and 26.6 mag (AB) 
respectively (Giavalisco et al. 2004).  At redshifts $z \sim 2$ and $z \sim 3$ these bandpasses collectively probe
rest-frame UV emission in the wavelength intervals $\sim 1500 - 3000$\,\AA\ and $\sim 1000 - 2000$\,\AA\ respectively.

\begin{figure*}
\epsscale{.9}
\plotone{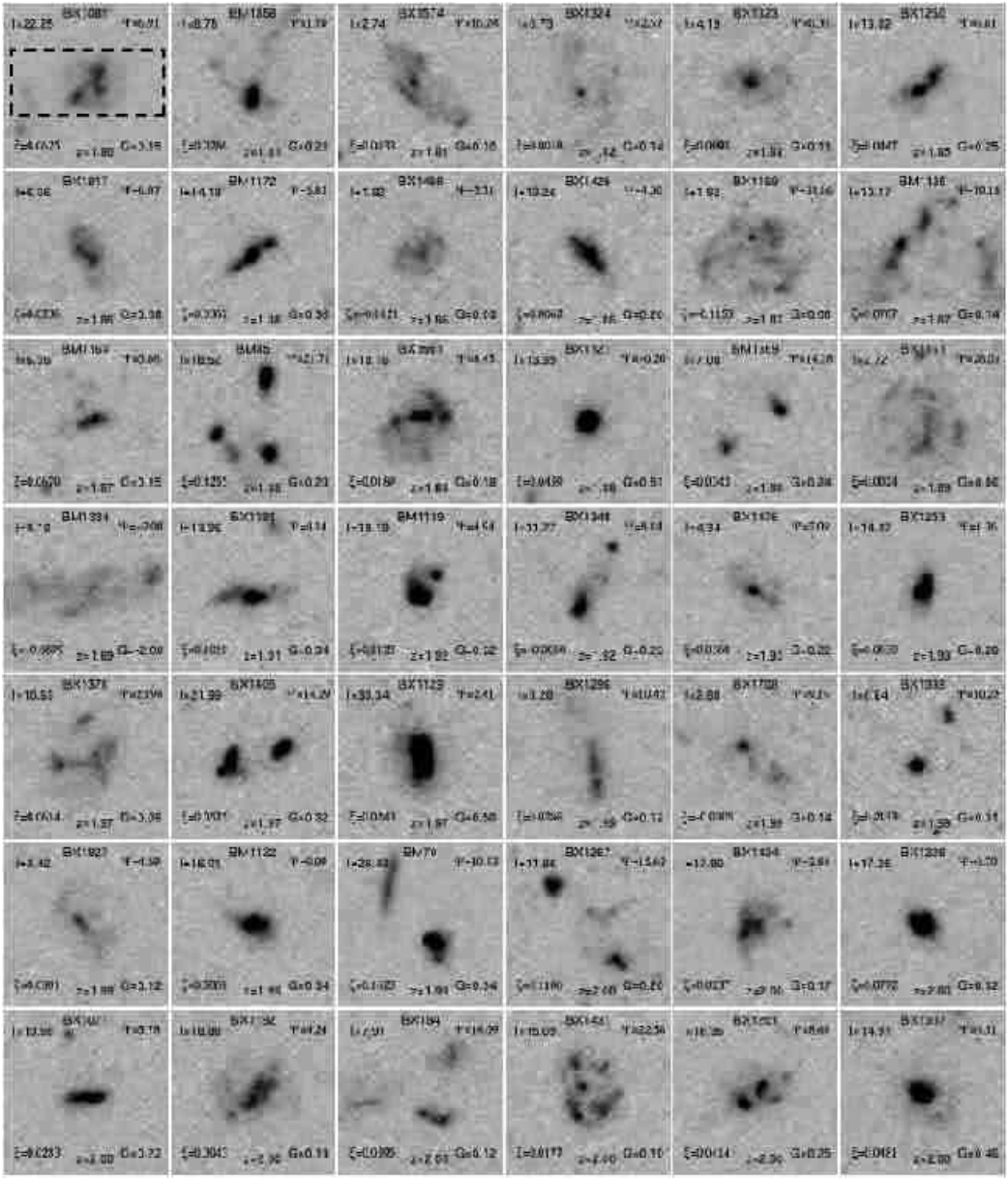}
\caption{{\it HST}-ACS rest-UV morphologies of the $z \sim 2-3$ galaxy sample, sorted in order of increasing redshift.
All panels include the galaxy name, redshift $z$, and morphological parameters size ($I$), gini ($G$), multiplicity ($\Psi$) and color dispersion ($\xi$).
Images are 3 arcseconds on a side, oriented with North up and East to the left.  Values of -2.00 for a particular morphological parameter 
indicate that a galaxy had too few pixels of suitably high surface brightness to define the parameter.  Greyscale is logarithmic and chosen to emphasize the visibility
of fainter nebulous regions, the details of high surface brightness features are thus suppressed.  The $1\farcs2$ width of the LRIS slit is indicated by
a dashed box in the top left panel.}
\label{mosaicA.fig}
\end{figure*}

The observed morphology is qualitatively similar throughout this range of wavelengths 
(see \S 3.6), and we therefore improve our signal-to-noise ratios per pixel by
creating a single rest-frame UV image from a weighted sum of the four individual bandpasses.  
Weights for this sum are determined proportionally to the inverse variance of the overall sky noise relative to the average number of counts from
the $z \sim 2-3$ galaxies.
The UV composite morphologies of our $216$ galaxies are shown in Figure \ref{mosaicA.fig} in order of increasing redshift and demonstrate
a variety of morphological types
ranging from single nucleated\footnote{We adopt the term ``nucleation'' to qualitatively describe a concentrated region of flux which might
naively be described as the ``nucleus'' of a given galaxy.  This contrasts with the term ``nebulosity'' which we use to describe diffuse flux
which is spread fairly uniformly over a number of pixels.} sources to extremely asymmetric 
sources with multiple nucleations and/or nebulous components.  The ``typical'' galaxy has a
morphology comprising one or more spatially distinct clumps with some degree of diffuse nebulosity, reminiscent of the HST-{\it STIS} UV morphology
of the local interacting galaxy VV~114 (Goldader et al. 2002) which is dominated by a patchy distribution of star formation regions.
Our initial morphological classification groups galaxies by visual inspection on the basis of the apparent 
nucleation of their light profiles and the presence and number of multiple nucleated emission components.  Galaxies fall within
five general classes:
\begin{enumerate}
\item Single strongly nucleated sources (11 sources at $z \sim 2$, 9 sources at $z \sim 3$).
\item Multiple strongly nucleated sources (6 sources at $z \sim 2$, 2 sources at $z \sim 3$).
\item Single nucleated source accompanied by nebulosity (61 sources at $z \sim 2$, 27 sources at $z \sim 3$).
\item Multiple nucleated sources accompanied by nebulosity (35 sources at $z \sim 2$, 12 sources at $z \sim 3$).
\item Nebulous emission with no strong nucleation (37 sources at $z \sim 2$, 16 sources at $z \sim 3$).
\end{enumerate}

\begin{figure*}
\plotone{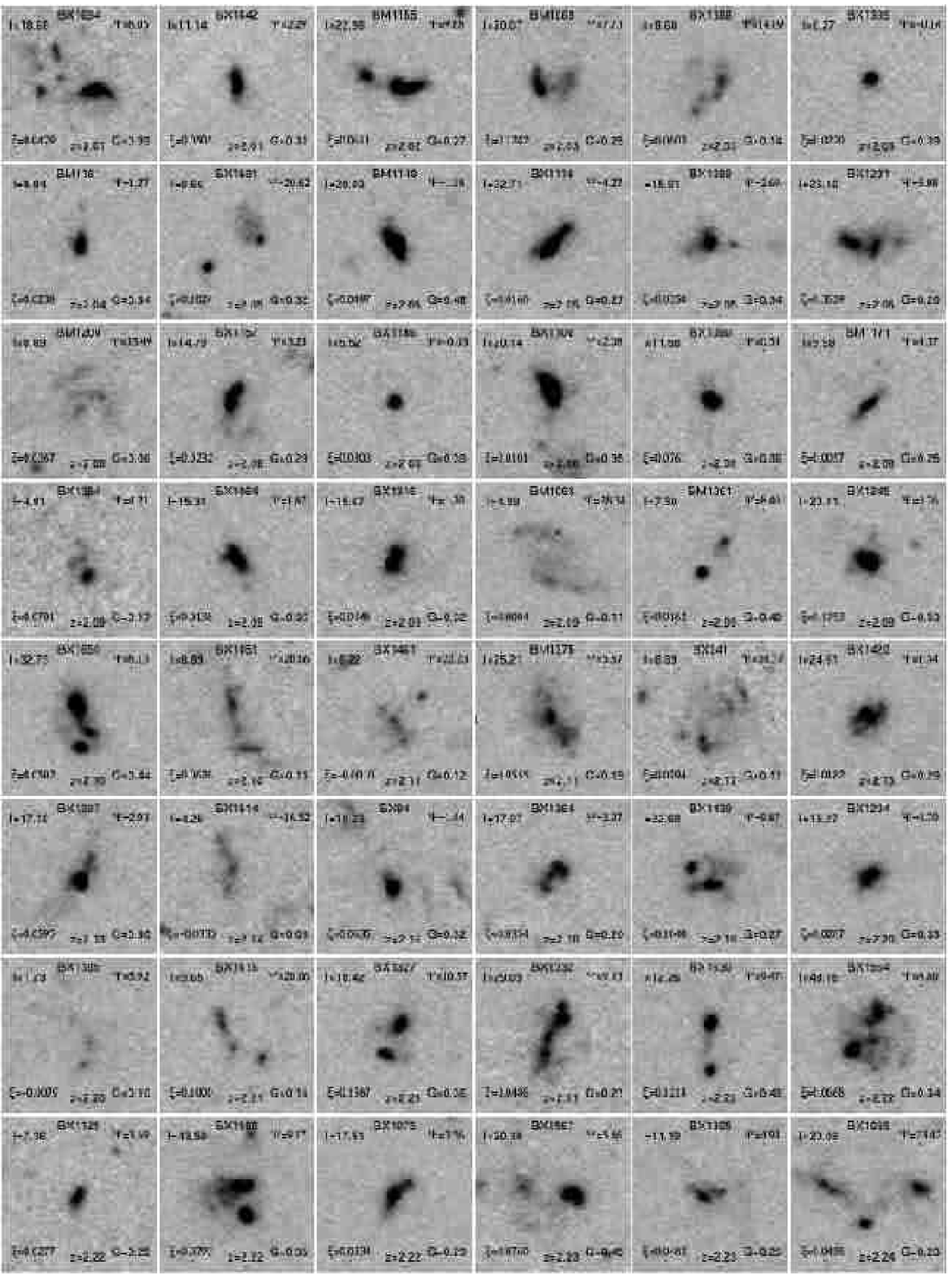}
\begin{center}
Figure~\ref{mosaicA.fig} (continued)
\end{center}
\end{figure*}

\begin{figure*}
\plotone{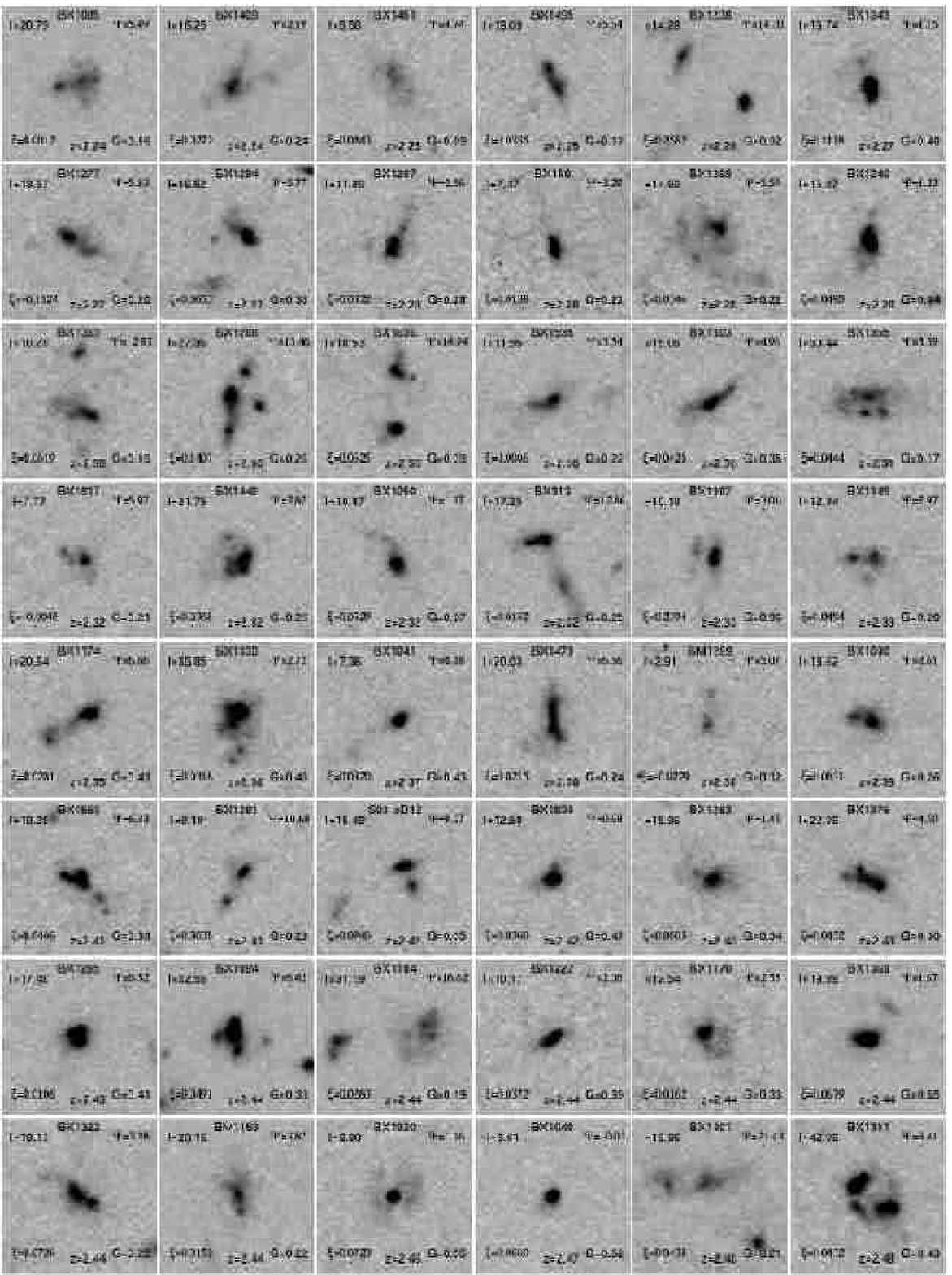}
\begin{center}
Figure~\ref{mosaicA.fig} (continued)
\end{center}
\end{figure*}

\begin{figure*}
\plotone{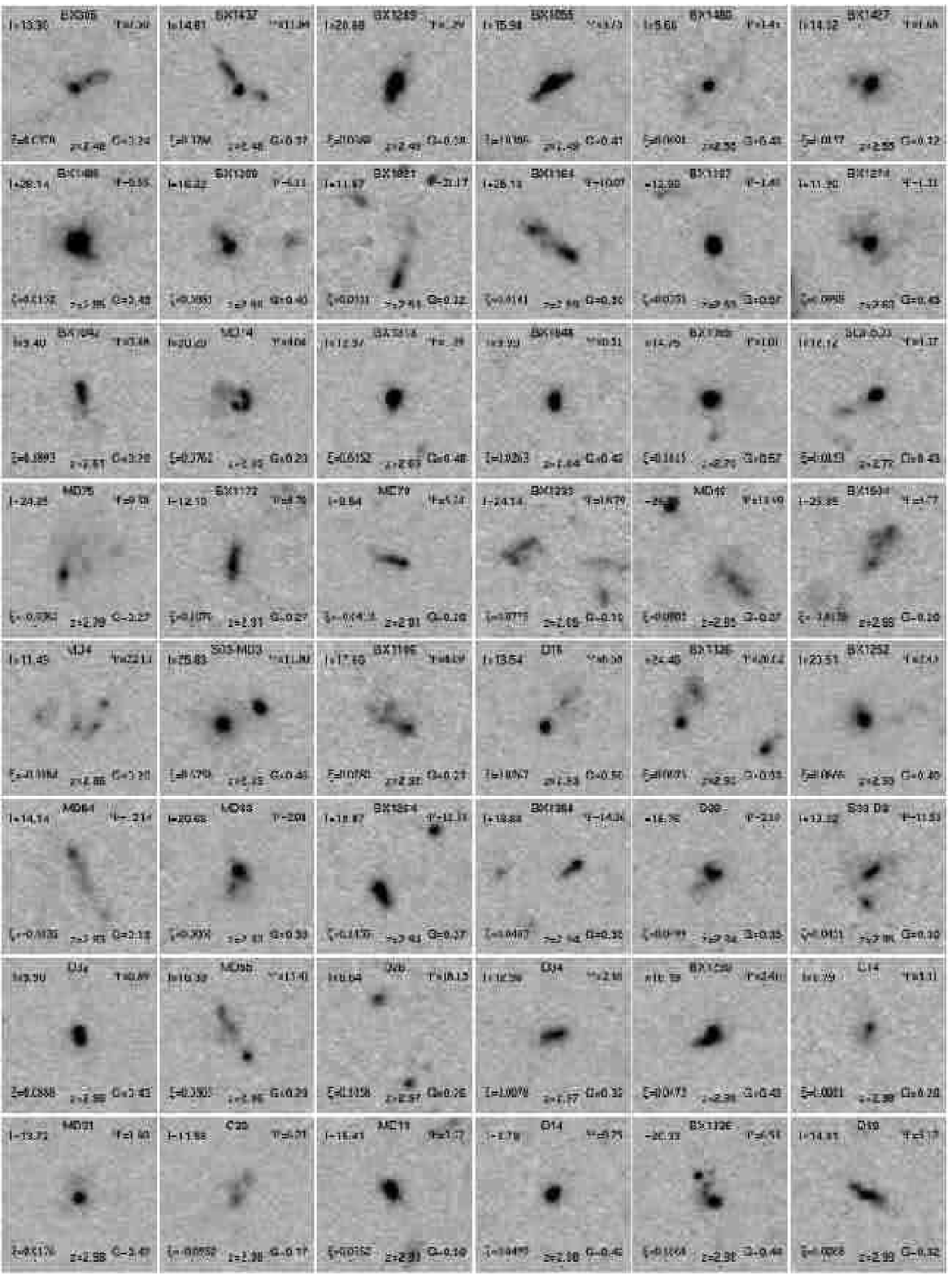}
\begin{center}
Figure~\ref{mosaicA.fig} (continued)
\end{center}
\end{figure*}

\begin{figure*}
\plotone{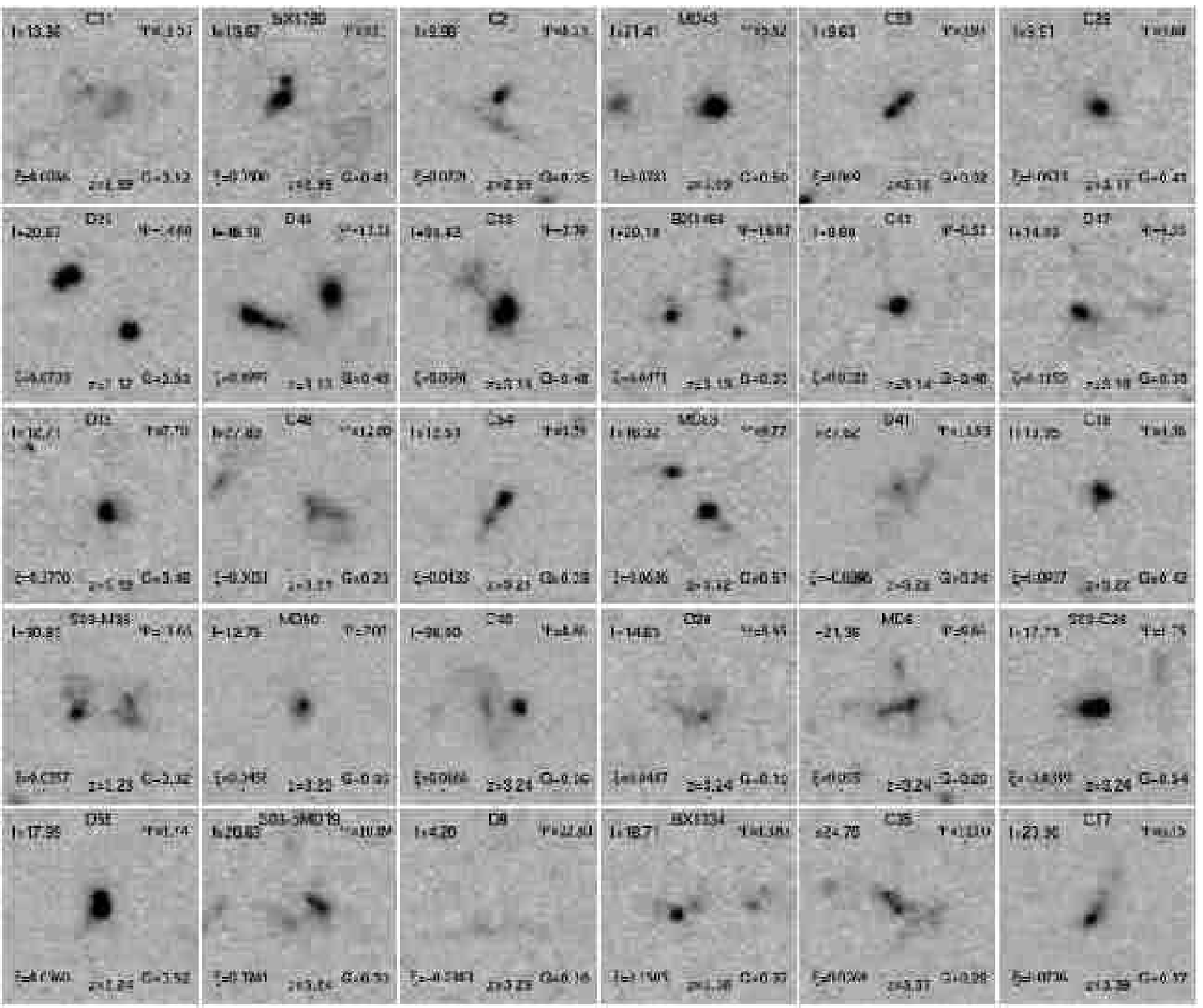}
\begin{center}
Figure~\ref{mosaicA.fig} (continued)
\end{center}
\end{figure*}

We seek a set of numerical parameters which will allow us to effectively reproduce these intuitive divisions,
while providing a more rigorous mathematical basis for the classification.
The ``$CAS$'' system of parameters has recently been a popular choice,
characterizing galaxies on the basis of their concentration ($C$; Kent 1985, Bershady et al. 2000), asymmetry ($A$, Schade et al. 1995),
and clumpiness ($S$, Conselice et al. 2003).  However, the first two of these quantities are explicitly defined with regard to circular or elliptical apertures
measured about a central point, which is only well defined
for galaxies with morphologies similar to traditional elliptical or spiral galaxies, while the third quantity relies upon suitable choice
of a smoothing scale on which clumpiness is defined.
In the case of the $z \sim 2 - 3$ galaxy sample, morphologies are generally so irregular (see Fig. \ref{mosaicA.fig})
that they do not have a well-defined ``center'', and the measured
values of the $CAS$ parameters can depend strongly upon the particular choice of center
\footnote{The asymmetry parameter $A$ is strictly found by numerically
searching through the image for the choice of center which minimizes the value of $A$
(see discussion by Conselice, Bershady, \& Jangren 2000).  While this relaxation technique partially
mitigates bias arising from poor centering, it does not address the underlying bias present in the assumption of circular symmetry
for galaxies as irregular as those depicted in Figure \ref{mosaicA.fig}.}
and smoothing length (see Lotz, Primack, \& Madau 2004, hereafter LPM04, for a detailed discussion).

We therefore favor a non-parametric approach to classification similar to that discussed by Abraham et al. (2003), LPM04,
and Lotz et al.  (2006) who define the gini coefficient $G$ as a measure of the uniformity of the flux distribution within a source.
In the following sections, we describe this and three additional non-parametric coefficients which we find effectively characterize the
irregular morphologies of these $z \sim 2 - 3$ galaxies.  We note that although we considered a 
host of additional parameters in our analyses (including the Petrosian radius and a non-parametric ``Petrosian area''),
we found that they provided no additional information and therefore omit them from further discussion.

\subsection{Pixel Selection}
It is of critical importance when measuring the morphologies of faint and highly irregular galaxies to apply uniform 
selection criteria by which to assign pixels to a galaxy as opposed to the surrounding sky
(i.e. defining the ``segmentation map'' of the source).
A variety of criteria have been adopted in previous studies, ranging from complex methods based upon curve-of-growth analysis (e.g. LPM04)
to basic surface brightness selection (e.g. Abraham et al. 2003).

The first of these methods, while robust to cosmological surface brightness dimming, can be non-trivial to implement in a manner consistent with non-parametric analysis.
As outlined by LPM04, the curve-of-growth method
calculates the elliptical Petrosian radius of a source (i.e. the radius from the center of the source
at which the average flux falls to a fixed fraction of the total inscribed flux; Petrosian 1976), and assigns to the segmentation map
all pixels within a suitably large radius of this center whose fluxes are greater than the value at the Petrosian radius.
Unfortunately, such a segmentation map enforces elliptical Petrosian
radii about a particular center and introduces biases similar to that of the $CAS$ system into the resulting morphological coefficients.  While such a
segmentation map is useful
for sources with approximately elliptical isophotes, we find that it tends to fail for galaxies with multiple components or extremely
irregular shapes since pixels at a particular ``Petrosian'' radius from an artificial center tend to include a large number of sky pixels,
decreasing the threshhold for surface brightness selection and resulting in some fraction of sky pixels being allocated to
the galaxy in the final segmentation map.  While the most noticeable cases may be fixed by hand, this
nonetheless introduces a bias as a function of morphological irregularity.  We explore
the effect of this bias on the measured gini coefficient in further detail in \S 3.3.

In contrast, the second of these methods (basic surface brightness selection) takes no account of surface brightness dimming but is more amenable to
non-parametric analysis.
However, with the aid of our confirmed spectroscopic redshifts for each galaxy in the optically-selected sample, we adapt
this morphology-independent
surface brightness selection technique to utilize a variable
threshhold tuned to select pixels in an identical range of intrinsic surface brightnesses at each redshift.

Our segmentation map is calculated as follows:  For each galaxy, we use our initial estimates of the position 
(based on seeing-limited $U_nG{\cal R}$ imaging)
to calculate the first order moment of the {\it HST}-ACS UV flux distribution within a 1.5 arcsecond (30 pixel) radius.  A revised value 
for the center is calculated
using this first order moment, and all pixels within a 1.5 arcsecond radius (i.e. slightly larger than the size of the largest
galaxy in our sample, so the exact position of the ``center'' is unimportant) 
of this new center are considered as possible candidates for assignment to the segmentation map.
While the ACS data product images have already been sky subtracted, we find that this subtraction is sometimes imperfect
and therefore subtract off residual sky flux measured in an annulus of radius 1.5 - 2 arcseconds around the revised center
(using a $3\sigma$ rejection algorithm to eliminate possible contaminating flux from sources which are nearby in projection).
Generally, these residual sky fluxes were small compared to the calculated object flux.

\begin{figure}
\plotone{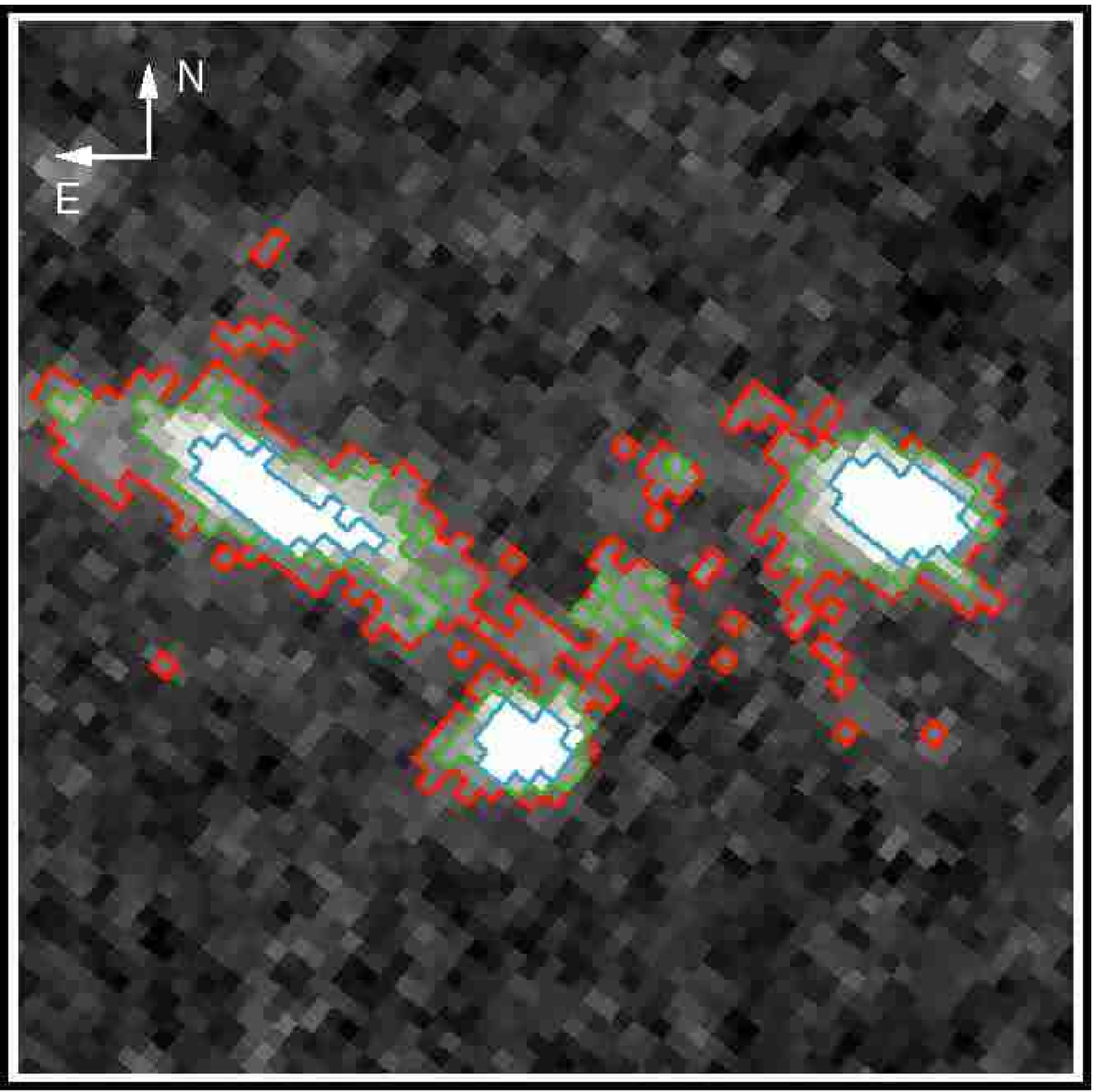}
\caption{Stacked ACS rest-UV 
image of BX 1035 ($z = 2.238$), overplotted with outlines of the 3, 5, and 10$\sigma$ segmentation maps (red, green, and cyan outlines respectively).  
The greyscale is linear in flux with the whitepoint set to $10\sigma$, the field of view is $3 \times 3$ arcseconds (corresponding to a physical region
$25 \times 25$ kpc at the redshift of the source).}
\label{sigmacuts.fig}
\end{figure}

Once this pre-processing is complete, we assign to the segmentation map all pixels whose 
flux is at least $n \sigma$, where $\sigma$ is the standard deviation of pixel values in the sky annulus, and where 
$n$ varies with redshift as
\begin{equation}
n = 3 \left(\frac{1+z}{1+z_{\rm max}}\right)^{-3}
\end{equation}
This variable surface brightness selection compensates for the effects of cosmological dimming throughout our range of sample
redshifts since (for a fixed observed bandpass) surface
brightness scales as $(1 + z)^{-3}$ and our selection criteria
therefore include pixels of the same intrinsic surface brightness at all redshifts (we make the assumption that there is no
intrinsic evolution of surface brightness with redshift, although see discussion in \S 3.2).  We set
$z_{\rm max} = 3.4$ (i.e. the upper end of our redshift distribution), so the value of the selection threshhold 
varies by a factor of about four across our redshift interval from $3\sigma$ at redshift $z = 3.4$
to $\sim 12\sigma$ at $z = 1.8$.  We neglect the change in angular size with redshift since
the angular diameter distance changes by only $\sim 13$ \% over $z \sim 1.8 - 3.4$ for our assumed cosmology.
The physical interpretation of our adopted segmentation map is illustrated in
Figure \ref{sigmacuts.fig}, which shows the pixels selected for a typical source according to 3, 5, and 10$\sigma$ criteria.

We note that by considering only flux within a 1.5 arcsecond radius we effectively restrict our sensitivity to morphology on distance
scales $\lesssim 13$ kpc at the redshift of our sample, similar to previous analyses (Conselice et al. 2003a, LPM04, Lotz et al. 2006)
whose selection radii range up to about 10 kpc.  Based on visual analysis of ACS and ground-based color maps, 
this appears to be the optimal choice of distance scale to include the majority of likely components for a particular galaxy whilst almost entirely excluding
probable contaminants which appear nearby in projection.
While it is likely that some gravitationally interacting systems extend considerably
beyond 13 kpc, we neglect these distant components in our characterization of the system morphology since
1) Our photometric and spectroscopic data is sensitive only to light at radii $\lesssim$ 1.5 arcseconds, and
2) Close interactions are more likely than distant to produce observable changes in the physical state of a galaxy.

\subsection{The Size Parameter: $I$}
The simplest morphological parameter to define is the projected physical size of the source ($I$) seen above our surface brightness threshhold 
(we use $I$ instead of the more intuitive $S$ to avoid confusion with the ``clumpiness'' parameter of Conselice et al. 2003).
Since the radius of a galaxy is not meaningful for multi-component systems, we use our spectroscopic redshifts to define $I$
as the total projected galaxy area in square physical kpc\footnote{Given the small change in angular diameter distance across
the redshift range of the sample there is little practical difference between using physical and angular sizes in our analyses.}.  That is,
\begin{equation}
I = N (0.05 \frac{{\rm arcsec}}{\rm pixel})^2 (2.4 \times 10^{-11} \frac{\rm ster}{{\rm arcsec}^2}) D_A^2
\end{equation}
where $N$ is the total number of 50 mas $\times$ 50 mas pixels in the segmentation map, and $D_A$ is the angular diameter distance
(in kiloparsecs) to
redshift $z$ for the assumed cosmology.
This parameter makes no attempt to discriminate between sources on the basis of the total amount or relative distribution of flux within the
segmentation map, and may therefore classify similarly both small strongly
nucleated sources (which have a low number of high flux pixels) and large yet extremely faint and nebulous sources (which have a low number of low flux
pixels barely satisfying the surface brightness selection criteria).

\begin{figure}
\plotone{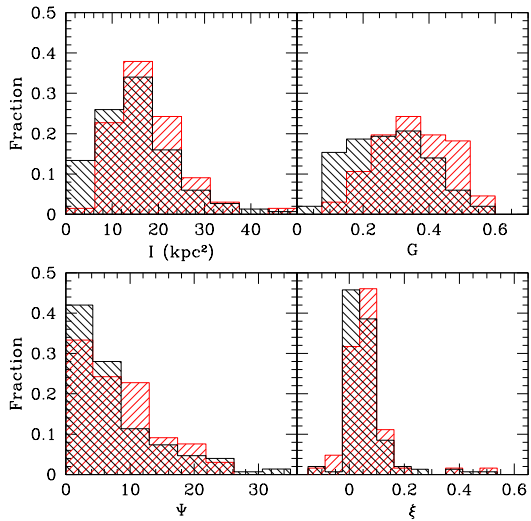}
\caption{Relative distribution of the morphological parameters $I$ (size), $G$ (gini), $\Psi$ (multiplicity), and $\xi$ (color dispersion) 
for the redshift $z \sim 2$ and $z \sim 3$ samples
(black and red histrograms respectively).  Counts are normalized by the total number of galaxies in each sample.}
\label{z.gmn.fig}
\end{figure}

In Figure \ref{z.gmn.fig} we plot the distribution of $I$ for the $z \sim 2$ and $z \sim 3$ samples (upper left panel, 
black and red histograms respectively).
$I$ typically ranges in value from less than 5 kpc$^2$ for faint and nebulous sources to around 10 kpc$^2$ for the 
most strongly nucleated single sources,
and up to as much as 48 kpc$^2$ for the brightest extended sources.
In the $z \sim 2$ and $z \sim 3$ samples
the mean projected size is $\bar{I} = 15.0 \pm 0.7$ kpc$^2$ and $\bar{I} = 17.4 \pm 0.9$ kpc$^2$ respectively\footnote{Uncertainties
quoted henceforth are uncertainties in the mean unless otherwise specified.}.
This small ($\sim 3\sigma$) apparent increase in average size from $z \sim 2$ to $z \sim 3$ 
arises because the $z \sim 3$ sample does not extend as far down the luminosity function as the $z \sim 2$
and small faint sources are therefore
underrepresented in the $z \sim 3$ sample.
If we compare only the fraction of the $z \sim 2$ sample that overlaps the $z \sim 3$ in rest-frame UV luminosity 
($L_{1600} > 2 \times 10^{10} L_{\odot}$; Reddy et al. 2006a)\footnote{62\% of the $z \sim 2$ sample have well determined measurements of $L_{1600}$, compared with
41\% of the $z \sim 3$ sample.  84\% of the $z \sim 2$ sample with measured $L_{1600}$ have $L_{1600} > 2 \times 10^{10} L_{\odot}$.}
the average value of $I$ at $z \sim 2$ rises to $\bar{I} = 16.7 \pm 1.0$ kpc$^2$, which is within $1\sigma$ of the $z \sim 3$ result.
Indeed, we note that in a recent analysis designed to more precisely measure the size evolution of galaxies
(as determined from their SExtractor half-light radii) Ferguson et al. (2004) found that the angular size of single-component
galaxies decreases with increasing redshift above $z \sim 1$.

\subsection{The Gini Parameter: $G$}
The gini coefficient ($G$, originally attributed to Corrado Gini [1912] and first introduced into the astronomical literature by Abraham et al. 2003) 
measures the cumulative distribution function of a population and may be used to distinguish between the cases for which flux is strongly nucleated versus
uniformly nebulous (see LPM04 for a detailed introduction).
Mathematically, $G$ may be calculated as (Glasser 1962):
\begin{equation}
G = \frac{1}{\bar{X} N (N-1)}\sum_{i=1}^N(2i -N -1)X_i
\end{equation}
where $\bar{X}$ is the average flux and
the $X_i$ pixel fluxes are sorted in increasing order before the summation over all $N$ pixels in the segmentation map.  The normalization 
prefactor multiplying the summation
ensures that $G$ takes values from zero to one inclusive, 
where low values indicate a uniform flux distribution and high values a concentration of flux in a few pixels.

Practically, we find that $G$ differentiates clearly between galaxy morphologies based on the degree of nucleation of their UV emission- 
sources which appear very nebulous generally have
$G \lesssim 0.2$, while sources with strongly nucleated emission have $G \gtrsim 0.5$ (see Fig. \ref{mosaicA.fig}).  The wide range of galaxies 
with some combination of nucleated and nebulous emission components
fall in a continuous distribution between these values.
We note that this range of values ($0 < G \lesssim 0.6$) is 
substantially different from that calculated for $z \sim 2 - 4$ HDF-N Lyman
Break galaxies  (LBGs)
by Lotz et al. (2004, 2006), who found $0.4 < G < 0.7$.
While some discrepancy is to be expected with the LPM04 results
since these authors used rest-optical data from HST-NICMOS (although this difference should be minimal
in light of the similarity between rest-UV and rest-optical morphologies),
we should expect greater similarity to the findings of Lotz et al. (2006) who used HST-ACS rest-UV data similar to our own.
This difference in results appears to
arise primarily from the adopted pixel selection method;
we find that elliptical Petrosian selection methods similar to that of LPM04 tend to include more sky pixels
at the ``Petrosian'' radius for highly irregular and nebulous objects (i.e. low $G$ in our sample), which lowers the threshhold for
assignment to the galaxy and includes more sky pixels in the resulting segmentation map.  
Including extra sky pixels in the map makes the genuine galaxy pixels appear to be comparatively more rich in flux, 
artificially increasing the value of $G$ calculated for the sources.
Applying a segmentation map similar to that of LPM04, we find that the gini coefficient calculated for our highest $G$ sources 
remains relatively unchanged, while our low gini sources increase their values of $G$
considerably, artificially compressing the range of values to $0.4 \lesssim G \lesssim 0.6$, closely mimicking the range of values presented
by LPM04.  We conclude that while the LPM04 values are accurate in the sense that a $G = 0.4$ (in their calculation) 
galaxy is more nebulous than a $G = 0.6$, the noise
introduced by this compression in dynamic range severely hampers the discriminating power of the gini coefficient.

A histogram of values of $G$ is plotted in Figure \ref{z.gmn.fig}, and suggests that
the mean value of the gini coefficient appears to change from $\bar{G}  = 0.27 \pm 0.02$ at $z \sim 2$
to $\bar{G} = 0.35 \pm 0.01$ at $z \sim 3$.  
To some extent this may be due to the underrepresentation of faint, nebulous objects in the $z \sim 3$ sample (see \S 3.2),
but even restricting the sample to the same range in intrinsic UV luminosity the difference between the two
samples is of order $3\sigma$ suggesting that the UV emission from objects at $z \sim 3$ is genuinely
slightly more nucleated than at $z \sim 2$, consistant with the finding of Ferguson et al. (2004) who demonstrated that apparent galaxy size
at constant luminosity decreases with increasing redshift.

We note that to first order, $G$ does not distinguish between sources based upon their number of nucleated components- i.e. a 
galaxy with two or more apparent nucleations has a value of $G$ nearly identical
to a galaxy with only one, so long as the cumulative distribution of light is similar.  That is, the exact {\it spatial} 
distribution of flux is irrelevant to $G$,
which is instead sensitive to the overall {\it curve of growth} of the total flux.

\subsection{The Multiplicity Parameter: $\Psi$} 

Our third classification parameter ($\Psi$) is designed to discriminate between sources based on how many apparent components
the light distribution is broken into, i.e. how ``multiple'' the source appears.
This parameter is similar to both the asymmetry parameter $A$ (Schade et al. 1995) and the second order moment of the 20\% brightest pixels ($M_{20}$,
LPM04) in that it is sensitive to the presence and distribution of
multiple clumps of flux.  However, both $A$ and $M_{20}$ have their limitations: $A$ depends strongly on the assumption of 
overall circular symmetry about some central point for each galaxy,
while $M_{20}$ is normalized by the moment of the segmentation map to remove the effect of overall galaxy size, which unfortunately results
in a limited dynamic range since the segmentation map and the 20\% brightest pixels often have a similar spatial distribution.
In contrast, $\Psi$ is defined in a manner that requires neither a center of symmetry nor a conventional normalization.

Using the observed flux distribution as a proxy for ``mass'', we calculate the ``potential energy'' of the light
distribution projected into our line of sight as
\begin{equation}
\psi_{\rm actual} = \sum_{i=1}^{N}\sum_{j=1, j\ne i}^{N}\frac{X_i X_j}{r_{ij}}
\end{equation}
where the summation of the pixel fluxes $X_i$ and $X_j$ runs over all $N$ pixels in the segmentation map, and $r_{ij}$ is the distance
(in pixels) between pixels $i$ and $j$.
This value is normalized by that which would be achieved with the most compact possible 
re-arrangement of the flux pixels, i.e. the configuration
which would require the most ``work'' to pull apart.  We re-arrange the physical positions of the $N$ 
pixels of the segmentation map
in a circular configuration with
the brightest pixel in the center and with pixel flux decreasing outwards with radius.  Calling the distance between pixels $i$ and $j$
in this re-arranged map $r'_{ij}$, the projected potential energy of this compact light distribution is
\begin{equation}
\psi_{\rm compact} = \sum_{i=1}^{N}\sum_{j=1, j\ne i}^{N}\frac{X_i X_j}{r'_{ij}}
\end{equation}

The multiplicity coefficient is then defined logarithmically as the ratio of these two quantities:
\begin{equation}
\Psi = 100 \times \textrm{log}_{10} \left(\frac{\psi_{\rm compact}}{\psi_{\rm actual}}\right)
\end{equation}

\begin{figure}
\plotone{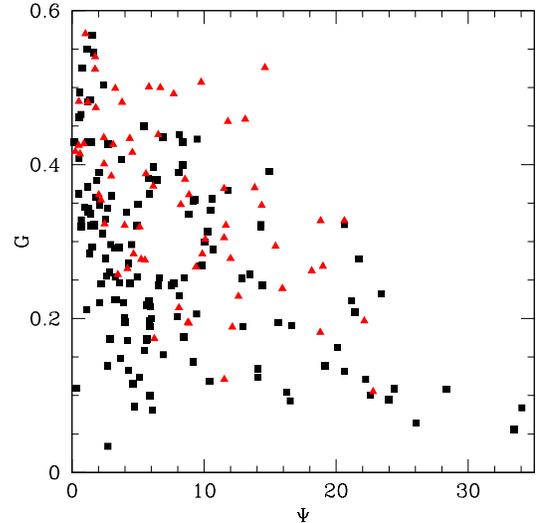}
\caption{Distribution of $G$ (gini) versus $\Psi$ (multiplicity) for galaxies in the $z \sim 2$ (black squares) and $z \sim 3$ (red triangles) samples.
Values are correlated in the sense that faint, nebulous galaxies tend to be broken into a greater number of non-contiguous components.}
\label{giniwork.fig}
\end{figure}

As shown in Figure \ref{z.gmn.fig}, $\Psi$ ranges from 0 up to about 30.
Typically, $\Psi = 0 -2$ for single well-nucleated sources, 
$\Psi \sim 5$ for sources which are beginning to show a second component in addition to the main nucleation,
and $\Psi \sim 10$ for strong double-nucleation systems.  At values of $\Psi > 10$, $\Psi$ increases as
the number and separation distance of nucleations increase, but
numerical distinctions become less obvious until at $\Psi \gtrsim 25$ the majority of sources are low $G$ nebulous sources for which
$\Psi$ breaks down as a useful statistic (see \S 3.6).
There is therefore some degree of correlation between $\Psi$ and the gini coefficient $G$
(Fig. \ref{giniwork.fig}) 
since sources with very nebulous emission (i.e. small $G$) tend to have very spread out flux distributions (i.e. large $\Psi$).  However,
this is of secondary importance to the largely orthogonal classification permitted by these two parameters;
while $G$ is most sensitive to the overall curve of growth of the flux distribution and fairly
insensitive to the number of nucleated components, $\Psi$ is sensitive to the number of bright components and comparatively insensitive to the
overall curve of growth.
We note that since $\Psi$ is a flux-weighted statistic it is more sensitive to bright pixels, and therefore a bright central source with an extremely
faint secondary source will tend to have low $\Psi$, while the same central source with a bright secondary source will tend to have higher $\Psi$.

\subsection{The Color Dispersion: $\xi$}

Finally, we capitalize on the available multi-wavelength {\it HST}-ACS data by defining
the color dispersion parameter $\xi$ (Papovich et al. 2003), which quantifies the degree of morphological difference between two bandpasses.
Applied between rest-optical and UV wavelengths, $\xi$ can measure the difference in spatial distribution between stellar populations
of differing ages, convolved with variations in the dust distribution and resulting extinction (Papovich et al. 2003).  
Using rest-UV data alone $\xi$ is a less powerful statistic, but nevertheless potentially informative.

We calculate $\xi$ as
\begin{equation}
\xi(I_V,I_z) = \frac{\sum(I_z - \alpha I_V - \beta)^2-\sum(B_z - \alpha B_V)^2}{\sum(I_z-\beta)^2 - \sum(B_z - \alpha B_V)^2}
\end{equation}
where $I_V, I_z$ are the object pixel fluxes in $V$ and $z$ bandpasses respectively\footnote{While 
use of the $B$ and $z$ bands would provide a greater wavlength baseline for morphological
differences, we use $V$ instead since $B$ is blanketed by absorption from the Lyman $\alpha$ forest for galaxies in the upper end of our
redshift range.}, $B_V,B_z$ the background sky flux in the bandpasses,
$\alpha$ is the flux ratio between the two bands, and $\beta$
represents the different in background levels between the two bands.
In brief, the first term in the numerator represents the summed square difference in pixel fluxes between the bands, the first term in the denominator
the summed square total object flux (for normalization), and the second terms in both numerator and denominator represent corrections
to statistically eliminate contributions to $\xi$ from the natural background sky variance.
Further details regarding the definition of this statistic are given by Papovich et al. (2003).

\begin{figure}
\plotone{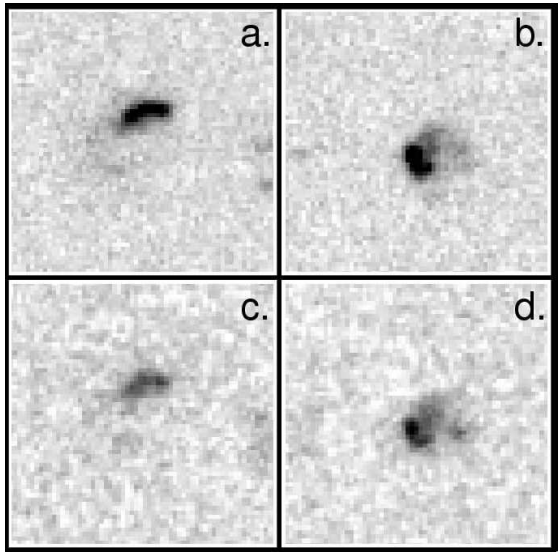}
\caption{HST-ACS images of BX 1157 and BM 1139 in F606W (Panels a and b respectively) and in F850LP (Panels c and d).  Both galaxies have
comparable redshifts ($z = 2.08 $ and 1.92), gini ($G = 0.291$, 0.321), and multiplicity ($\Psi = 3.23$, 4.94) in the summed rest-UV image.
Both galaxies show similar decrease in signal-to-noise ratio from F606W to F850LP, but have widely variant color dispersions ($\xi = 0.023, 0.112$).
Image size and orientation are the same as in Figure \ref{mosaicA.fig}.}
\label{bandcomp.fig}
\end{figure}

As shown in Figure \ref{z.gmn.fig}, typical values range from $\xi \sim 0.0$ to $0.15$, with $\bar{\xi} = 0.057 \pm 0.006$ at $z \sim 2$
and $\bar{\xi} =  0.061 \pm 0.011$ at $z \sim 3$ (i.e. consistent with a constant value to well within the uncertainty).
We note that this range of values is larger than that of $\xi \sim 0.0 - 0.05$
found by Papovich et al. (2005) for a sample of $z \sim 2.3$ galaxies measured between rest-UV and optical bandpasses.
The origin of this discrepancy is uncertain, although it is likely that different galaxy samples, segmentation maps, and
bandpasses all contribute.
By and large, it appears that our calculated $\xi$ is dominated mainly by scatter rather than by genuine differences between the
apparent morphology in $V$ and $z$ bands.  Figure \ref{bandcomp.fig} shows that for two galaxies with $\xi = 0.023$ and $\xi = 0.112$
the most obvious trend from $V$ to $z$ bands is an overall decrease in the signal-to-noise ratio.
Given the similarity of the effect in both cases, it is not readily apparent why the two galaxies
should have such different color dispersions.

\begin{figure}
\plotone{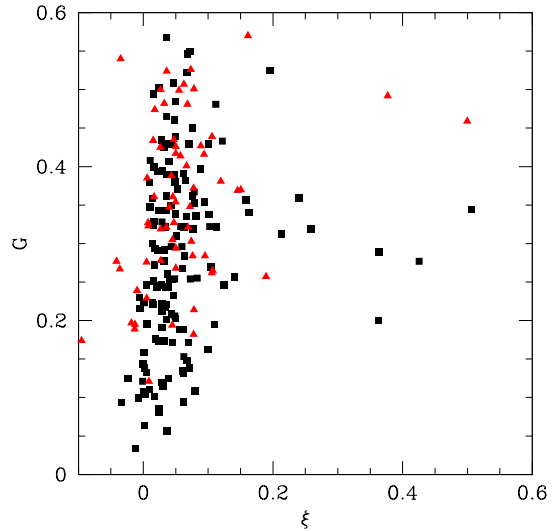}
\caption{Distribution of $G$ (gini) versus $\xi$ (color dispersion) for galaxies in the $z \sim 2$ (black squares) 
and $z \sim 3$ (red triangles) samples.
Note that a few extreme outliers from this trend lie outside the visible region of this plot to better illustrate the main correlation.}
\label{ginixi.fig}
\end{figure}

Curiously, while $\xi$ has no obvious correlation with a visible difference in morphology between bands, it does correlate with the Gini coefficient
$G$ of a source (as shown in Fig. \ref{ginixi.fig}) in the sense that $\xi$ is (on average) slightly larger
in the most nucleated sources.  This is likely a consequence of
the flux-weighting of $\xi$- high flux pixels have greater absolute variation between bandpasses and tend to dominate the sum
in Equation 7, increasing $\xi$ for high $G$ sources where $\xi$ is dominated by variation from a few bright pixels rather than
averaging over a larger number of lower-variation pixels for low $G$ sources.
In conclusion, we caution that while $\xi$ is ideally a useful statistic, in the present case it may be too erratic to provide a
great deal of information.

\subsection{Robustness of the Parameters}

\begin{figure*}
\plotone{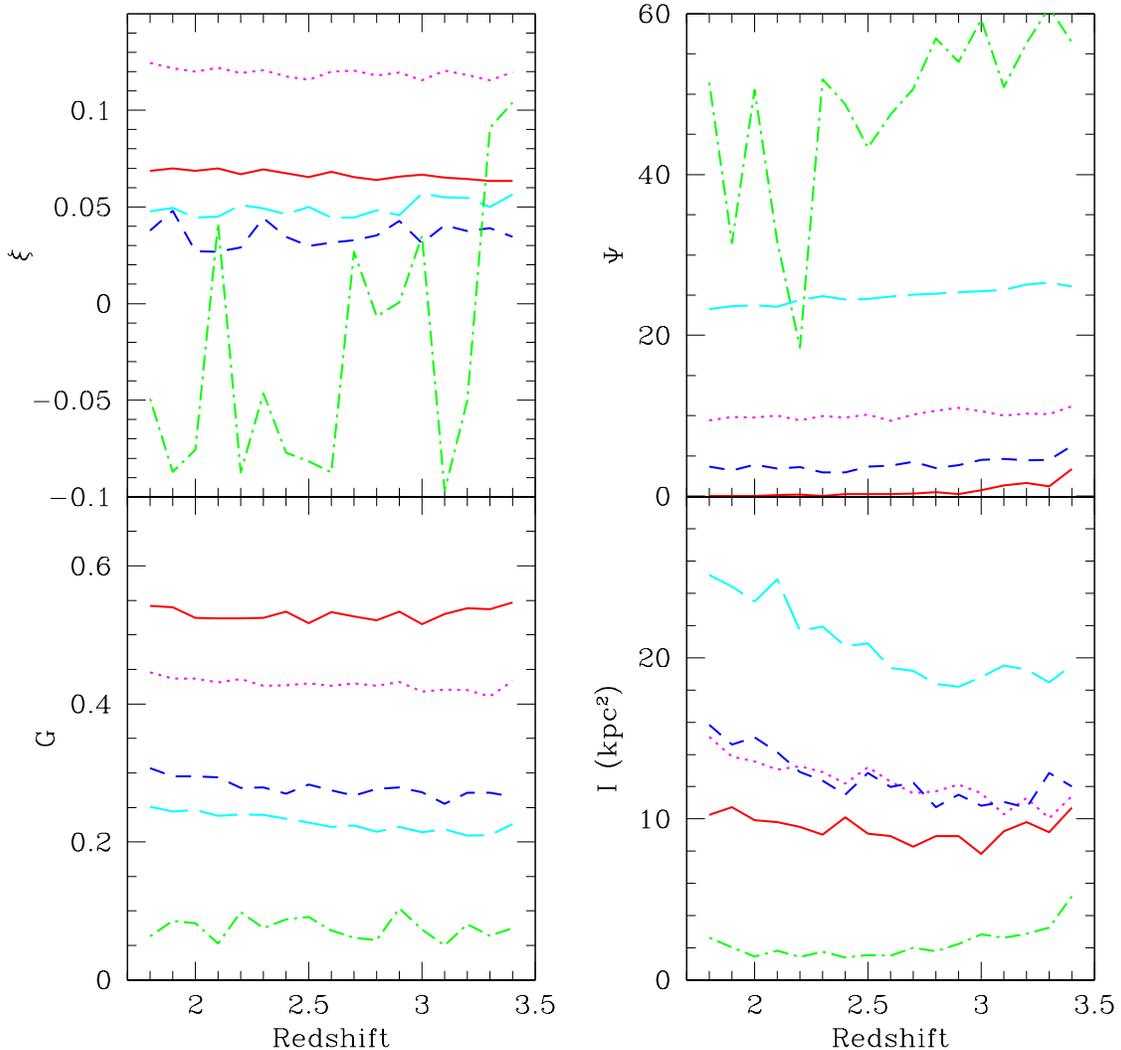}
\caption{Figure illustrates the robustness of morphological parameters to redshift when using the variable surface brightness method to define the segmentation map.
Colored lines indicate the trends found by artificially redshifting one representative galaxy from each of the five general morphological classes through the
range $z = 1.8 - 3.4$ before performing morphological analysis.  These five classes (defined in \S 3) and their representative galaxies are: 
Single nucleated source (BX 1040: solid red line), multiple nucleated source (BX 1630: dotted magenta line), single nucleated source with nebulosity (BX 1297: 
short-dashed dark blue line), multiple nucleated source with nebulosity (BX 1035: long-dashed cyan line), and purely nebulous source (BX 1169: dot-dashed green line).}
\label{ztrend.fig}
\end{figure*}

We test the robustness of these parameters to cosmological distance by selecting one representative galaxy for each of the five general morphological
categories (defined previously in this section), artificially redshifting them through the range $z = 1.8 - 3.4$, and measuring the resulting
morphologies using our variable threshhold pixel selection technique.  As illustrated by Figure \ref{ztrend.fig}, 
in most cases we would measure consistent values for the morphological parameters for a given galaxy if it were located at any redshift throughout our
sample.  This constancy fails however for the most faint and 
nebulous of objects ($G \lesssim 0.15$, green dot-dashed line in Figure \ref{ztrend.fig}), for which the multiplicity $\Psi$
and the color dispersion $\xi$ can vary subtantially because the number of pixels in the segmentation map is small and random
variations in noise can drastically affect both the locations of selected pixels (to which $\Psi$ is particularly sensitive) and the residual color
(to which $\xi$ is sensitive).  Such objects represent less than 20\% of the sample however, and the importance of this effect on $\Psi$ may be mitigated
by noting that $\Psi$ is only well-defined up to $\Psi \sim 25$.
We also note a slight decrease in $I$ with redshift to $z = 3$ , followed by a smaller upturn at $z > 3.3$.  The first of these effects
is consistent with our neglect of the change in angular size with redshift, and the second with the inclusion of a small number of sky pixels in the segmentation
map when the selection threshhold is pushed down to 3$\sigma$.  Neither effect is large enough to noticeably impact our analysis.

\begin{figure*}
\plotone{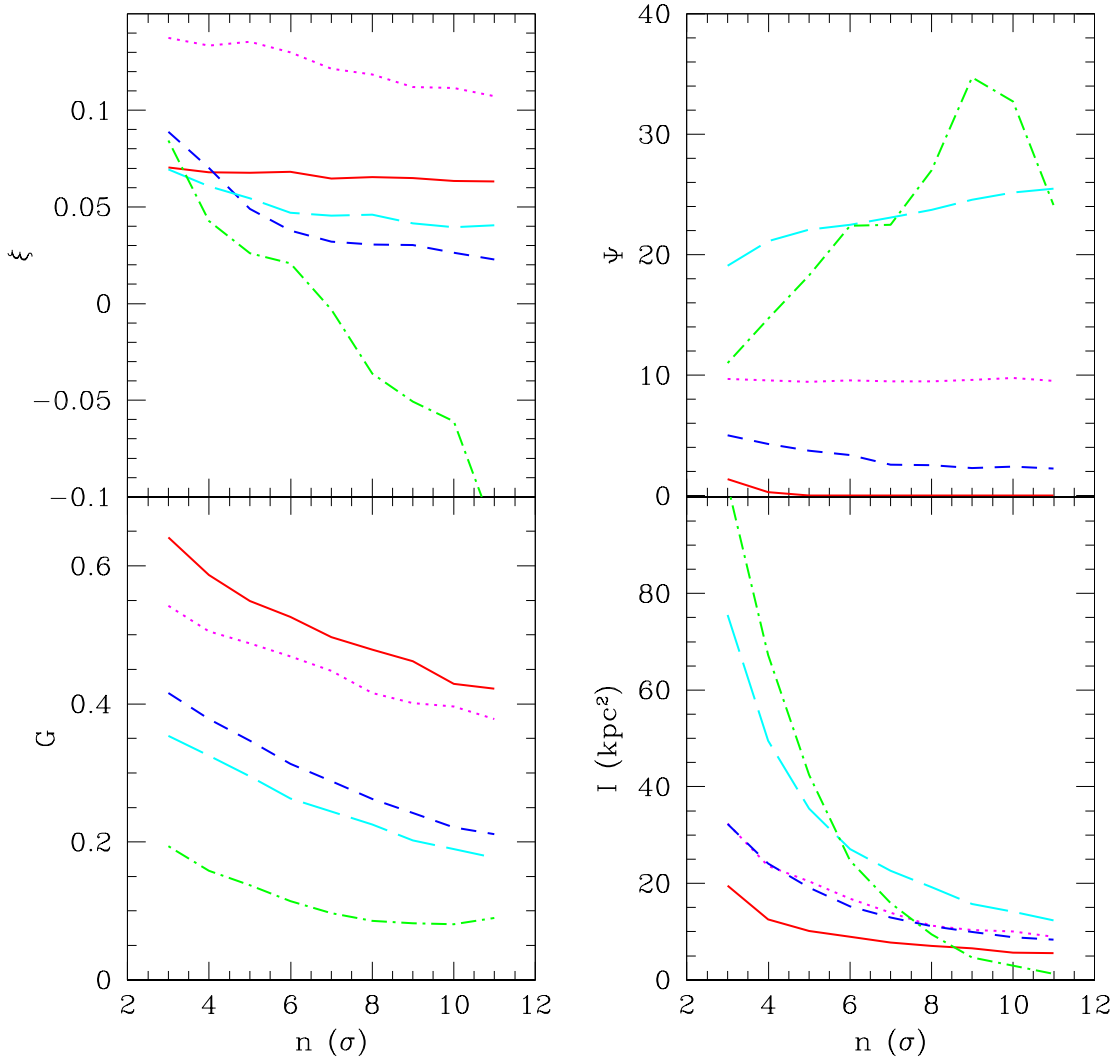}
\caption{As Figure \ref{ztrend.fig}, but plots the calculated values of morphological parameters for the five classes of galaxies as a function of the adopted
surface brightness selection threshhold in multiples of the background sky noise $\sigma$.  
Note for comparison that segmentation maps corresponding to the 3, 5, and 10$\sigma$ selection regions for BX1035 (i.e. long-dashed
cyan line) are shown in Figure \ref{sigmacuts.fig}.}
\label{sbtrend.fig}
\end{figure*}

Having demonstrated the uniformity of the selection technique, we consider what (if any) information is lost for galaxies in the lower end of the redshift range
by effectively restricting our analysis to those pixels brighter than about $10\sigma$ above the background sky noise.  In Figure \ref{sbtrend.fig} we plot
morphological parameters calculated for the same five galaxies as before for a range of surface brightness selection threshholds.  Most obviously, 
$I$ decreases rapidly with increasing brightness threshhold since correspondingly fewer pixels are included in the segmentation map.
In addition, $G$ declines noticeably as the lower-flux population is gradually
omitted from the map.
In contrast, $\Psi$ and $\xi$ remain relatively constant throughout the range of selection threshholds considered, with the exception
of purely nebulous sources (dot-dashed green line) for which we have previously noted the instability of these two parameters.
Most importantly however, if we neglect purely nebulous sources we note that the same {\it relative} information is preserved at all brightness threshholds-
no matter what the threshhold it is equally possible to distinguish between the galaxy types despite the overall trends.  That is, since the lines do not
cross (again with the exception of the green dot-dashed line and $\xi$ at very low threshholds) we may be confident that we are not discarding
information by using a $\sim 10\sigma$ segmentation map at lower redshifts.

We now note a few caveats to this analysis which should be borne in mind.
First, by using a fixed bandpass we have measured the morphologies of $z \sim 2$ and $z \sim 3$ galaxies at slightly different rest wavelengths.  
However, this effect should be negligible since our
morphological parameters do not change substantially using data from the $B$, $V$, or $I$ bandpasses (although they can change somewhat in the $z$ bandpass due
to the lower signal-to-noise ratio for all detections at near-IR wavelengths).  Parameters calculated in each of these bands correlate very strongly with one
another, differences being dominated by the limiting surface brightness reached in each band.
Given that our sample consists of actively star-forming galaxies, this is perhaps not surprising since young massive stars tend to dominate
the SED out to at least rest-frame blue wavelengths.
Second, we have neglected surface brightness evolution throughout our sample.  
Since our primary goal is to investigate the physical meaning of morphology
rather than providing a detailed comparison between samples at different redshifts we choose not to introduce uncertain corrections for such effects
into our analysis.
Third, it is possible that some regions of flux included in our segmentation maps are due to chance alignments of some object
at a very different redshift than the target galaxy.  However, we judge this to be an unlikely major source of contamination to our sample
since 1) the various pieces of the galaxies have consistent colors, and 2) the total number density of optically selected $z \sim 2 - 3$ galaxies is
$\sim$ 10 arcmin$^{-2}$ to ${\cal R} = 25.5$ (Steidel et al. 2004), giving a low probability for chance supposition 
of two unrelated galaxies within $1.5$ arcseconds.
Finally, we have made no correction for the effects of galaxy inclination to the line of sight.
While it is conceivable, and indeed likely, that some of the variance in apparent morphology may be due simply to different orientations
relative to the line of sight, our knowledge of these galaxies is at present insufficient to allow us to compensate for such effects
in any meaningful way.

\begin{figure*}
\plotone{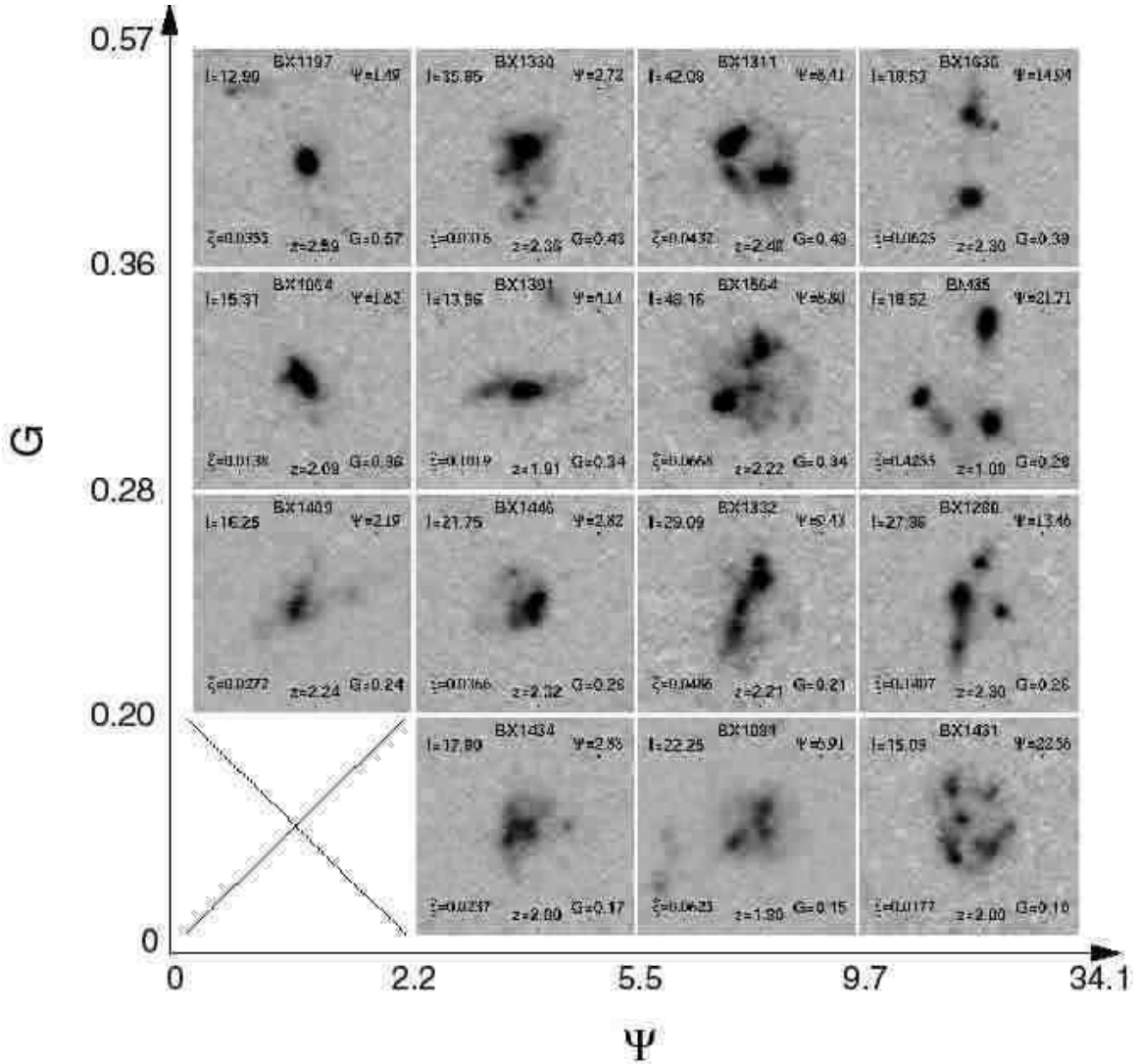}
\caption{{As Figure \ref{mosaicA.fig}: \it HST}-ACS rest-UV morphologies of redshift $z \sim 2$
galaxies classified according to gini ($G$) and multiplicity ($\Psi$) parameters.  Horizontal and vertical bin ranges are chosen to
divide the sample into quartiles.  Increasing values of $G$ correspond to increasing nucleation of source
emission, and increasing values of $\Psi$ correspond to increasing number of components.}
\label{gallery.fig}
\end{figure*}

Of the four parameters which we have introduced, the $G$---$\Psi$ classification scheme best reproduces the morphological trends apparent to the eye.
This system is illustrated in Figure \ref{gallery.fig}, which shows a representative set of
galaxies from our sample and demonstrates how these two parameters serve to distinguish
galaxies based on their degree of nucleation and number of components.  
Since surface brightness dimming and bandshifting make it difficult to associate the morphologies of
$z \sim 2-3$ sources with galaxies in the local universe we avoid such direct, and potentially misleading comparisons.
We describe the {\it sense} of our classification parameters by noting, however, that early-type galaxies typically are larger (i.e. larger $I$), more
concentrated (larger $G$), less multiple (lower $\Psi$), and have lower color dispersion ($\xi$) than their late-type counterparts which are instead dominated by emission
from multiple scattered knots of star formation.  A more thorough overview of the relation between
local and high-redshift galaxies is given by LPM04.

\section{OVERVIEW OF REST-UV SPECTRA}
\subsection{Spectral Processing}
As part of an ongoing Keck LRIS-B spectroscopic survey (Steidel et al. 2004) we have compiled UV spectra of 
our 216 target galaxies in the GOODS-N field, and
here explore how these spectra correspond to the morphology of their host galaxies.  Unfortunately, spectra of individual galaxies are rarely
of sufficient quality to accurately measure the strengths of their emission/absorption components, and we therefore divide the 
spectra into five bins according to each of our four morphological
parameters and measure the strengths of features in stacks of the spectra within these bins.
We find that five bins gives a suitably large number of bins across which
spectroscopic trends may be assessed while still producing reasonable quality stacked spectra (each comprised of 30 galaxies at $z \sim 2$ and 13 galaxies
at $z \sim 3$).

Our spectroscopic combination method is similar to that described by Shapley et al. (2003).
Before stacking, individual sky-subtracted spectra were flux-calibrated and shifted to the systemic rest frame using
the prescriptions given by Adelberger et al. (2005), then resampled to a common dispersion of 1\,\AA\ pixel$^{-1}$ and
rescaled to a common mode in the range $\lambda\lambda 1250-1500$\,\AA.  Spectra were manually cropped to eliminate overly noisy segments due to
falloff of the LRIS-B blue-side efficiency and of the red-side dichroic transmission.
The resulting spectra were averaged
using a min/max rejection of hot pixels, cosmic rays, and bad sky subtraction events; three high and low values were rejected at each dispersion coordinate (corresponding
to a rejection of the high/low 10\% for the $z \sim 2$ sample).
Continuum normalization was performed by iteratively fitting a spline function to the stacked spectrum using wavelength intervals 
selected to be free of strong interstellar features, giving results consistent with the normalization method described by Rix et al. (2004).


\subsection{Key Spectral Features.}
\begin{figure*}
\epsscale{0.65}
\plotone{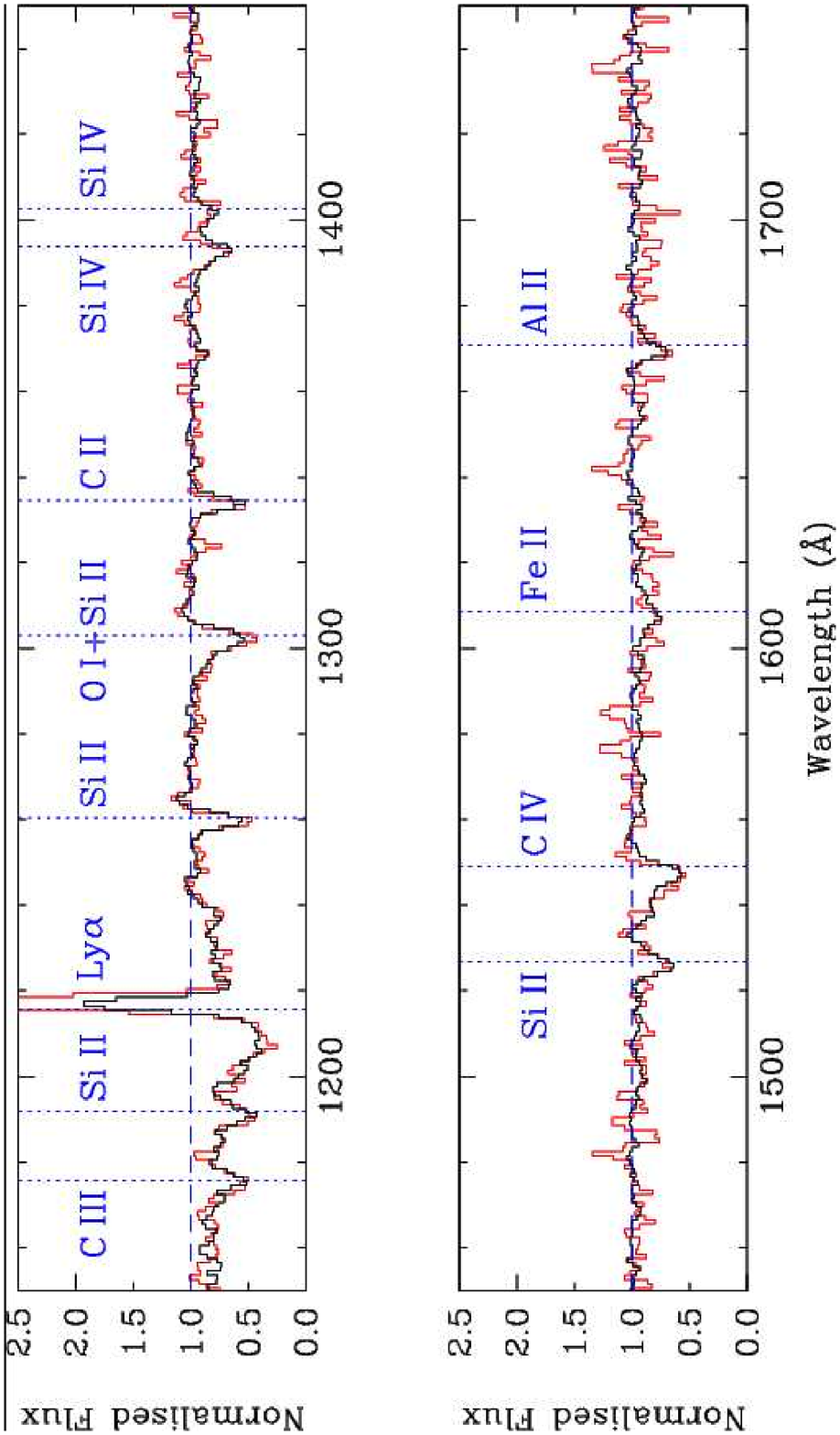}
\caption{Stacked continuum-normalized spectra are plotted for all 150 $z \sim 2$ and 66 $z \sim 3$ galaxies (black/red lines respectively).  
Labels and dotted lines indicate the fiducial locations of atomic transitions corresponding
to major spectroscopic features.  
Note that Ly$\alpha$ emission is considerably stronger in the $z \sim 3$ sample than the $z \sim 2$, and peaks at a relative normalized 
flux of 3.95 (not shown for clarity).}
\label{zspec.fig}
\end{figure*}

The rest-frame ultraviolet ($\lambda\lambda 1000-1800$\,\AA) 
spectra of rest-UV color selected galaxies are typified (Fig. \ref{zspec.fig}) by redshifted Lyman $\alpha$ (Ly$\alpha$) emission 
superposed on a broad blueshifted resonant absorption trough, and by
strong velocity-broadened absorption lines due to energetic galaxy-scale outflows (Shapley et al. 2003).
The strength of Ly$\alpha$ emission is originally governed by the rate of star formation and initial mass function (IMF) in galactic H\,{\sc ii} regions, although
this raw flux is significantly modified by resonant scattering from interstellar H\,{\sc i}, generally culminating in its eventual absorption by interstellar
dust and subsequent reradiation in the infrared.  The observed Ly$\alpha$ emission strength is therefore a complex function of
neutral hydrogen column density, dust fraction, and geometric/kinematic projection of the outflowing ISM on the line of sight.
Typically, observed Ly$\alpha$ emission is redshifted by roughly 400 km s$^{-1}$ relative to the systemic velocity
(Shapley et al. 2003, Steidel et al. 2004), corresponding to scattering from the back
side of outflowing ISM, from which the Doppler shift is suitable to take the photon off of resonance and permit it to escape the galaxy.

Other notable features include strong, optically thick absorption lines from low-ionization species 
(e.g. Si\,{\sc ii}~$\lambda1192$, Si\,{\sc ii}~$\lambda1260$, O\,{\sc i}+S\,{\sc ii}~$\lambda1303$, 
C\,{\sc ii}~$\lambda 1334$,
Si\,{\sc ii}~$\lambda1526$, Fe\,{\sc ii}~$\lambda 1608$, 
and Al\,{\sc ii}~$\lambda1671$) arising from outflowing neutral ISM.  
As shown in Figure \ref{zspec.fig}, these absorption lines are
blueshifted from the systemic redshift, corresponding to absorption
from outflowing gas seen from the nearby side of a galaxy-wide outflow --- their full widths
can reach $\sim 1000$\,km~s$^{-1}$ (Pettini et al. 2002; Shapley et al. 2003).
We note that while
O\,{\sc i}+Si\,{\sc ii}~$\lambda 1303$ (a blend of the low ionization 
species O\,{\sc i}~$\lambda 1302$ and Si\,{\sc ii}~$\lambda 1304$) appears to be 
blended with a third component around $\lambda 1297$
(resulting in a shallower blue-side slope for the composite line) it is generally possible to deblend this additional component to measure
O\,{\sc i}+Si\,{\sc ii}~$\lambda 1303$  alone.
This broad $\lambda 1297$\,\AA\ feature is likely itself a blend of the stellar photospheric lines C\,{\sc iii}~$\lambda 1296.33$, 
Si\,{\sc iii}~$\lambda 1294.54$, Si\,{\sc iii}~$\lambda 1296.73$,
and Si\,{\sc iii}~$\lambda 1298.93$ (Tremonti et al. 2006), 
variations in the relative strengths of which can shift the apparent centroid of the $\lambda 1297$\,\AA\ blend
from $\lambda = 1295$ to 1298\,\AA.

Also apparent are absorption lines due to higher-ionization species, including the 
Si\,{\sc iv}~$\lambda\lambda 1393, 1402$ doublet 
and C\,{\sc iv} absorption around $\lambda 1549$\,\AA.  Although all of these lines 
are blended with stellar features, C\,{\sc iv} is a particularly complex blend of interstellar absorption lines at $\lambda 1548$\,\AA\ and 1550\,\AA, 
combined with a P-Cygni component from the winds of the most luminous O and B stars.
We neglect the numerous additional features due to nebular, fine-structure, and stellar atmosphere transitions
(e.g. C\,{\sc iii}~$\lambda 1176$; see Shapley et al. 2003 for a further list), 
since they are not generally detected to high statistical significance after the
sample has been divided amongst five morphological bins.
We refer the reader to Shapley et al. (2003) and references therein for further discussion of the physical interpretation
of the rest-UV spectra.


\subsection{Equivalent Widths and Uncertainties}
The equivalent widths of all absorption features were integrated non-parametrically relative to the normalized continuum.
Generally, features shortward of Ly$\alpha$ and longward of $\lambda1500$ \AA\ are 
considerably noisier than those between these wavelengths, and we therefore define average
``low-ionization'' and ``high-ionization'' absorption line strengths ($W_{\rm LIS}$ and $W_{\rm HIS}$ respectively) as
the weighted means of the shorter wavelength transitions 
Si\,{\sc ii}~$\lambda1260$, 
O\,{\sc i}+Si\,{\sc ii}~$\lambda1303$, and 
C\,{\sc ii}~$\lambda 1334$ (for $W_{\rm LIS}$) 
and Si\,{\sc iv}~$\lambda 1393$\,\AA\ 
and $\lambda 1402$\,\AA\ (for  $W_{\rm HIS}$).

The total equivalent width of the Ly$\alpha$ feature (i.e. absorption plus emission) tends to be strongly affected by the noisiness
of the absorption trough, so we characterize the strength of the emission component alone. 
The equivalent width of emission ($W_{\rm Ly\alpha}$) is determined by dividing the total flux in the emission component 
by the continuum level which would be present in the absence of any absorption trough.

It is possible to assess the statistical significance of possible deviations from a constant value by using the $\chi^2$ statistic
\begin{equation}
\chi^2 = \sum_i \left(\frac{x_i - \bar{x}}{\sigma_{\rm sample}}\right)^2
\end{equation}
where $x_i$ is the measured equivalent width for a given bin, $\bar{x}$ is the mean equivalent width among all five bins, and $\sigma_{\rm sample}$
is the square root of the natural variance among samples of randomly binned spectra.\footnote{Thirty samples was sufficient to 
converge $\sigma_{\rm sample}$ to within 5\%.}
We find that $\sigma_{\rm sample}$ is considerably larger than the uncertainty in the mean equivalent width for a given stack (which belies the actual
variance observed between randomly-drawn samples), and therefore adopt it as a conservative means of ensuring that any possible trends
are more significant than would be likely to occur randomly.
As such, the significance of a particular value of $\chi^2$ is evaluated using $5-1=4$ degrees of freedom to give the likelihood $P$ that deviations of the five
measurements from a constant average value 
(whether in the form of a monotonic trend or a single bin whose equivalent width varies greatly from the average with respect to the expected variance)
are greater than that expected from a random binning of the sample.
Applying this method to our measurements from randomly selected quintiles of galaxies,
we determine that  a threshhold of $P > 90\%$ suffices to cull apparent associations due to random variance.

\section{THE RELATION OF REST-UV MORPHOLOGIES TO SPECTRA}

\begin{figure*}
\plotone{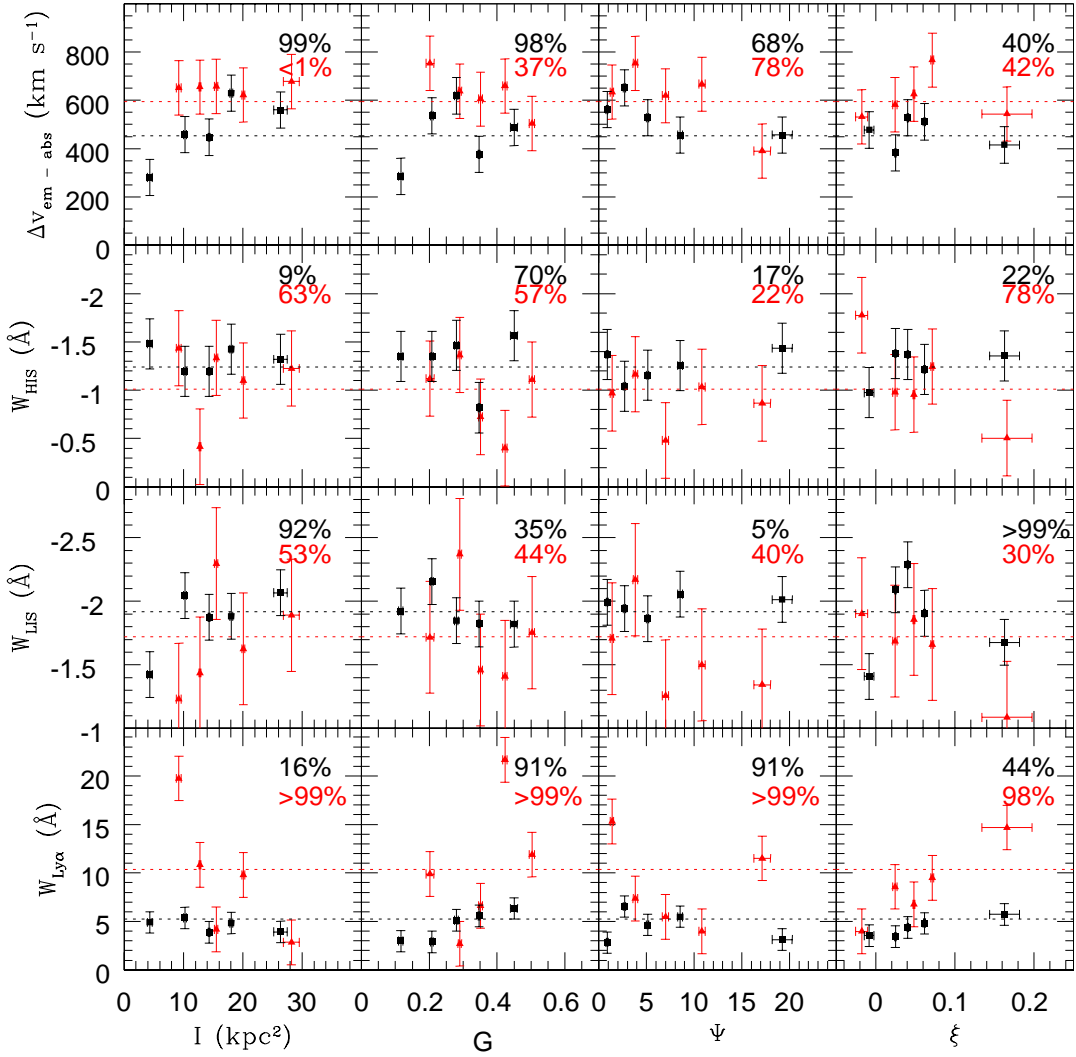}
\caption{Ly$\alpha$ emission ($W_{\rm Ly\alpha}$) equivalent width,
low- and high-ionization interstellar absorption equivalent width ($W_{\rm LIS}$
and $W_{\rm HIS}$ respectively), and the kinematic offsets between emission and absorption lines ($\Delta v_{\rm em - abs}$)
are shown measured from quintiles in the morphological parameters $I$ (size), $G$ (nucleation), $\Psi$ (multiplicity), and $\xi$ (color dispersion).  
Black and red points represent data
from the $z \sim 2$ and $z \sim 3$ samples respectively, shown with error bars representing the standard deviation among measurements made
from randomly binned samples.
Red and black dotted lines indicate the values measured for the complete $z \sim 2$ and $z \sim 3$ stacks respectively, while 
the black and red percentages given in each panel indicate the $\chi^2$ probability $P$ that the data points shown are statistically
inconsistent with a constant value.  Generally, values of $P > 90$\% are indicative of significant deviation- note particularly the linear trends of Ly$\alpha$
emission increasing for more nucleated sources (i.e. those with higher $G$) and low-ionization absorption strength increasing for larger sources
(i.e. those with greater $I$).}
\label{speccor.fig}
\end{figure*}

As described in \S 4.1, we bin the galaxy sample separately into quintiles according to each of the morphological parameters $I$, $G$, $\Psi$, and $\xi$.
Bin divisions are determined so that each bin contains an identical number of galaxies (i.e. 30 each for the $z \sim 2$ sample, and 13 each for 
the $z \sim 3$ sample), precise ranges are given in Table \ref{granges.table}.  
In all cases our spectra are broadly consistent with the diffuse light spectrum of local starburst galaxies 
(e.g. Chandar et al. 2005)
indicating that largely similar processes likely dominate the UV light output of all morphological types in our sample.
Despite this general similarity the spectra show some variation with morphology as demonstrated in Figure \ref{speccor.fig}, which
we proceed to discuss in detail.

We first note however the caveat that the position angles of the slits used to obtain
our UV spectra have not been chosen to correspond to the major axis
of each of the galaxies, and it is therefore possible for a few of the most widely separated sources with multiple components
that the UV spectra may represent only one of the components.  Given the seeing limited nature of the spectroscopic observations 
and the fact that the typical size of our
targets ($\sim$ 1 arcsecond) is less than the width of the LRIS-B slit ($1\farcs2$)
we doubt that this has a considerable effect.  More likely perhaps is the probability that spectra are dominated by light
from the bright nucleated regions of our sources and may not be expected to show any difference between nucleated and nucleated + nebulous sources
if the spectrum of the nucleated regions are similar in each case.

\begin{deluxetable}{ccccccc}
\tablecolumns{7}
\tablewidth{0pc}
\tabletypesize{\scriptsize}
\tablecaption{Quintile bins for morphological parameters.}
\tablehead{
\colhead{Parameter} & \colhead{$x_1$\tablenotemark{a}} & \colhead{$x_2$} & \colhead{$x_3$} & \colhead{$x_4$} & \colhead{$x_5$} & \colhead{$x_6$}}
\startdata
$I$ ($z \sim 2$) & 0.181 & 7.473 & 12.540 & 15.977 & 20.108 & 48.163\nl
$I$ ($z \sim 3$) & 4.200 & 11.559 & 13.952 & 17.991 & 21.407 & 45.181\nl
$G$ ($z \sim 2$) & 0.034 & 0.172 & 0.245 & 0.321 & 0.382 & 0.568\nl
$G$ ($z \sim 3$) & 0.105 & 0.265 & 0.321 & 0.385 & 0.459 & 0.570\nl
$\Psi$ ($z \sim 2$) & -0.199 & 1.667 & 3.871 & 6.151 & 10.677 & 34.062\nl
$\Psi$ ($z \sim 3$) & 0.249 & 2.426 & 5.236 & 8.849 & 12.599 & 22.795\nl
$\xi$ ($z \sim 2$) & -0.115 & 0.015 & 0.033 & 0.049 & 0.076 & 0.507\nl
$\xi$ ($z \sim 3$) & -0.095 & 0.008 & 0.040 & 0.057 & 0.078 & 0.500\nl
\enddata
\tablenotetext{a}{Successive values of $x_i$ denote boundary divisions between galaxies binned into
equal-size quintiles according to each of the morphological parameters (without regard to the other three parameters).}
\label{granges.table}
\end{deluxetable}

\subsection{Interstellar Absorption Lines}


As illustrated in Figure \ref{speccor.fig} (middle rows), 
we find that the strength of interstellar absorption lines ($W_{\rm LIS}$ and $W_{\rm HIS}$) 
is largely uncorrelated with UV morphology, although there appears to be a statistically
significant ($P = 92$\%) association of the strength of low-ionization species with galaxy size $I$.  Indeed, 
the data may be consistent with a trend
that larger galaxies tend to have stronger interstellar absorption lines.
Such a trend is most apparent for the $z \sim 2$ sample; while the $z \sim 3$ sample suggests a similar trend the greater statistical uncertainties
mitigate its significance.

\begin{figure}
\plotone{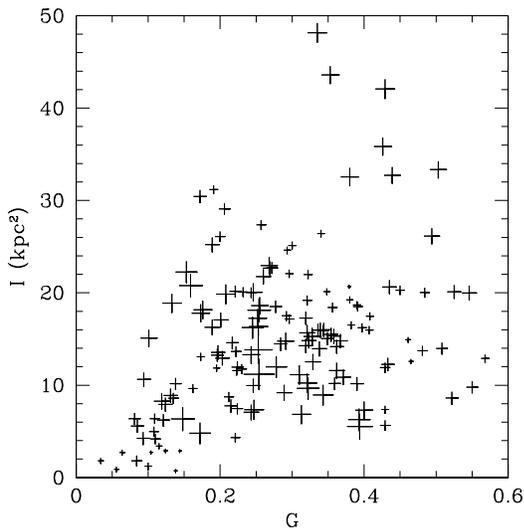}
\caption{Low-ionization absorption line strengths for $z \sim 2$ galaxies are plotted (plusses) according to their location in the size---gini ($I-G$) plane, spectra
of each galaxy have been stacked with those of their ten nearest neighbors in this plane.
Point sizes correspond linearly to the value of $W_{\rm LIS}$.  Note the overabundance of weak-absorption sources (smaller symbols) in the lower left corner of the plot, and in the strip running from $G,I$ = 0.2, 30 kpc$^2$ to 0.6, 10 kpc$^2$.}
\label{ginii.lis.fig}
\end{figure}

We explore this possible trend in greater detail for the $z \sim 2$ sample
by stacking our spectra in two dimensions to explore the variation of line strengths across
a plane of two parameters smoothed by a variable-width kernel.  We stack the spectrum of
each galaxy with that of its 10 nearest neighbors in the $I-G$ plane to distinguish large nebulous objects from large yet nucleated objects, 
with inter-point ``distances'' along each axis normalized by the dynamic range of values along the axis.
In Figure \ref{ginii.lis.fig} we plot $I$ versus $G$ for $z \sim 2$ galaxies, with point size corresponding to the strength of $W_{\rm LIS}$.
On the whole, the distribution of galaxies with strong and weak low ionization lines (i.e. large and small points) is quite similar, 
except for the lower left corner representing the faintest and most nebulous galaxies, for which there is an
overabundance of small points (i.e. sources with weak $W_{\rm LIS}$).  Given the similar overall
distribution of line strengths, 
we conclude that the association between $I$ and low-ionization absorption strength is due
to the over-representation of weaker-line sources among the faint and nebulous galaxy sample rather than to an overall trend.
Although this may indicate a genuine physical characteristic, we note that it is this class of low surface-brightness galaxies for which 
spectra are typically of the poorest quality and absorption line measurements least reliable.
It is interesting, however, to note that the high ionization absorption lines do not likewise appear weaker in this class of galaxies
(see Fig. \ref{speccor.fig}) as might be expected were the apparent decline in low-ionization line strength due to spectrum quality alone.

Figure \ref{ginii.lis.fig} also introduces another curiosity, namely the overabundance of weak-line objects in a line running through the
plot from $G,I = 0.2, 30 $ kpc$^2$ to $G,I = 0.6, 10 $ kpc$^2$.  This band does not correspond to an obviously distinct class of galaxies,
or have any obvious reason for occupying the region of this plot which it does.  We suggest the possibility that there could be a population
of galaxies which have relatively weak outflowing components which are somehow distinguishable in this plot, but caution that random variation
combined with the kernel smoothing technique might also be responsible for this enigmatic feature.

There is one additional significant deviation from uniformity for the color dispersion $\xi$ and the low-ionization absorption line strength
(at a significance level of $>99\%$ for the $z \sim 2$ sample).  However, inspection of Figure \ref{speccor.fig} offers no clear
explanation of the nature of this association: objects in both the smallest and largest bins of $\xi$ have weaker
absorption lines than objects in the intermediate three bins, and there hence appears to be no particular {\it linear} correlation between the variables.
While this deviation may be a genuine product of physical phenomena its interpretation is unclear, particularly given the uncertain meaning of $\xi$.

\subsection{Ly $\alpha$ Emission}

\begin{figure}
\plotone{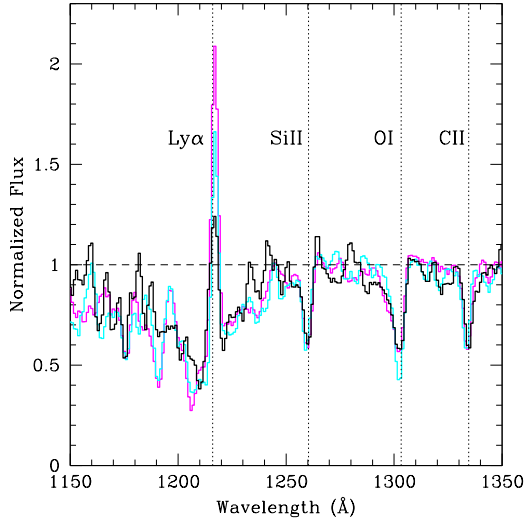}
\caption{Stacked, continuum-normalized 
spectra for $z \sim 2$ galaxies representing three bins in the morphological nucleation parameter $G$.  The low $G$ nebulous galaxy bin is represented by
the black line, the intermediate bin by the cyan line, and strongly nucleated sample by the magenta line.
Spectra have been smoothed by a three pixel boxcar filter and normalized consistent with the prescriptions of Rix et al. (2004).
Labels and dotted lines indicate the fiducial locations of major spectral features.}
\label{ginispectra.fig}
\end{figure}

As indicated by Figure \ref{speccor.fig} (bottom row) there are relatively many possible associations between Ly$\alpha$ emission
strength and galaxy morphology, the simplest of which to interpret is the positive correlation between Ly$\alpha$ and the gini coefficient $G$.
These two parameters show a clear linear trend of increasing emission strength with increasing nucleation; for the $z \sim 2$ sample the most
strongly nucleated sources show roughly twice the emission seen in the most nebulous 
(see also Fig. \ref{ginispectra.fig}), and for the $z \sim 3$ sample an even stronger trend is possible
(although less well defined).  
It is unlikely that this trend is due to substantial variation of the composition or velocity of interstellar gas since such variations
should also affect the strengths of absorption lines, which remain statistically constant across all values of $G$ (see Figs. \ref{speccor.fig} and
\ref{ginispectra.fig}).  A more plausible explanation may be 
that dust might be present in greater quantities in nebulous low-$G$ sources, causing greater attenuation
of Ly$\alpha$ photons.  Under this hypothesis,
it may be simply the presence of more or less dust which
determines both the observed degree of UV nucleation and the strength of resonant Ly$\alpha$ emission.  We explore this hypothesis further in \S 6.

\begin{figure}
\plotone{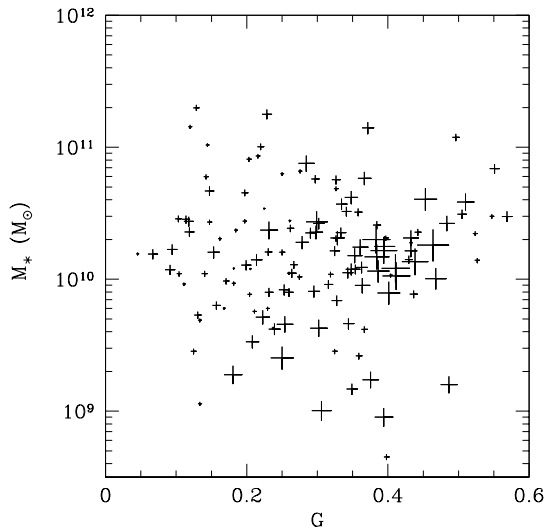}
\caption{Ly$\alpha$ emission line strengths for $z \sim 2$ galaxies are plotted (plusses) according to their location in the stellar mass---gini ($M_{\star}-G$) plane, spectra
have been stacked with their ten nearest neighbors in this plane as described in \S 5.1.
Point sizes correspond linearly to the value of $W_{\rm Ly\alpha}$.  Note the concentration of sources with strongest Ly$\alpha$
emission around $G \sim 0.4$ and  $M_{\star} \sim 1.5 \times 10^{10} M_{\odot}$.}
\label{ginimass.lya.fig}
\end{figure}

We map the Ly$\alpha$ trend in greater detail in Figure \ref{ginimass.lya.fig}, which compares the strength of the association between Ly$\alpha$
and $G$ with the previously known association between Ly$\alpha$ and stellar mass $M_{\star}$ (i.e. that galaxies with higher stellar masses have lower
average Ly$\alpha$ emission strength, Erb et al. 2006b)
\footnote{We caution that despite the fact that $G$ is correlated with $W_{\rm Ly\alpha}$ and $W_{\rm Ly\alpha}$
with $M_{\star}$, $G$ itself is {\it not} correlated with $M_{\star}$ (see \S 6), indicating that correlations
are not necessarily commutative.}.
Figure \ref{ginimass.lya.fig} suggests that the overall association between $G$ and Ly$\alpha$ is genuine, but that emission strength in fact peaks for
galaxies with $G \sim 0.4$ rather than for the few galaxies with $G > 0.5$.  Since the galaxies with $G > 0.5$ represent the extreme of nucleation and 
have much weaker Ly$\alpha$ emission than expected based on the majority of the galaxy sample we posit that these few galaxies may be somewhow
distinct from the rest of the sample.
Alternatively, the $G \sim 0.4$ galaxies may represent a particularly dust-free population from which it is possible to see both star forming regions
and their surrounding material.

We also observe also that $W_{\rm Ly\alpha}$ deviates from statistical uniformity when binned according to multiplicity $\Psi$ (for both $z \sim 2$ and $z \sim 3$
samples), apparent size $I$ (for the $z \sim 3$ sample), and color dispersion $\xi$ (for the $z \sim 3$ sample).
In the first case, there is less of an obvious trend in Figure \ref{speccor.fig}
than an apparently random scatter of points, with particularly discrepant points found in both the highest and lowest bins in $\Psi$.
In the latter two cases, $W_{\rm Ly\alpha}$ possibly declines with $I$ and increases with $\xi$ for the $z \sim 3$ sample, but these trends are absent
from the $z \sim 2$ spectra.
It is possible that these apparent deviations are due to the residual correlation between our morphological parameters
(in which case the sense of a trend would be a complicated projection
of the $G$---$W_{\rm Ly\alpha}$ correlation onto the density distribution in morphology space), or simply that these deviations are telling us of some
trend or population of galaxies which is not well distinguished using our chosen morphological parameters (or indeed any of the many others which we have explored).
However, we are inclined (despite our conservative estimates of the uncertainty) to ascribe at least those deviations only visible in the $z \sim 3$ spectra
to statistical variance since similar trends do not appear in the much higher-quality composite spectra from the $z \sim 2$ galaxy sample.

\subsection{Kinematic Offsets}

Since our stacked spectra have all been shifted to the systemic rest frame (i.e. that of the stars and [H\,{\sc ii}] regions), it is possible
to calculate the kinematic offset $\Delta v_{\rm em - abs}$ in each stack between Ly$\alpha$ emission and the average of the
low-ionization absorption lines 
Si\,{\sc ii}~$\lambda1260$, 
O\,{\sc i}+Si\,{\sc ii}~$\lambda1303$, and 
C\,{\sc ii}~$\lambda 1334$.
For the $z \sim 2$ and $z \sim 3$ samples respectively we find that
$\Delta v_{\rm em - abs} = 453$ and 594 km s$^{-1}$, roughly similar to the Shapley et al. (2003) value of 
$\Delta v_{\rm em - abs} = 510$ km s$^{-1}$ found for a large stack of 794 $z \sim 3$ LBG spectra
\footnote{We note that the average velocity offset in {\it individual} $z \sim 3$ LBG spectra with both emission and absorption components is typically
$\sim 650$ km s$^{-1}$- see Shapley et al. (2003) for details.}.
While our relatively small sample of 66 spectra at $z \sim 3$ is too noisy to distinguish any significant trends with morphology,
for the $z \sim 2$ sample $\Delta v_{\rm em - abs}$ is convincingly correlated with both the size parameter $I$ and the gini coefficient $G$
(see Fig. \ref{speccor.fig}, top panels).

In particular, the offset velocity increases nearly linearly from $\sim 300 $ km s$^{-1}$ for the smallest galaxies with $I \sim 5$ kpc$^2$ 
to near 600 km s$^{-1}$ for the largest with $I \sim 20$ kpc$^2$.  The differences are similarly pronounced with respect to the gini parameter $G$,
but instead of increasing linearly through the sample, Figure \ref{speccor.fig} suggests that $\Delta v_{\rm em - abs}$ is lowest for galaxies
with $G \lesssim 0.2$, peaks for intermediate values of $G \sim 0.3$, and possibly declines slightly at the most nucleated values of $G \gtrsim 0.5$.
This high $\Delta v_{\rm em - abs}$ sample of high $I$ and moderate - high $G$ corresponds reasonably well to a particular morphological sample-
the Type 4 galaxies (as identified in \S 3) which occupy a large angular area and tend to have bright nebulous emission paired
with one or more distinct nucleations.

These data suggest that larger, more UV luminous ($I$ correlates well with UV luminosity- see \S 6) galaxies on average
may have stronger outflows than the rest of the population, as might be expected if these galaxies have particularly energetic input to
their interstellar media and are therefore capable of blowing the most energetic outflows.  We caution however that the major discrepant point
in both of these trends is for the most nebulous sample (i.e. that with both small $I$ and small $G$) whose UV spectra are fainter and typically
of slightly lower quality.

Pairing this with the most significant understandable trend discovered in \S 5.1 and 5.2, namely the positive correlation between
$G$ and Ly$\alpha$ emission strength, we might be led to conclude that Ly$\alpha$ 
and $\Delta v_{\rm em - abs}$ are positively correlated, in constrast
to the results of Shapley et al. (2003) who found that 
outflow velocities are slightly weaker for $z \sim 3$ LBGs with strong Ly$\alpha$ emission.
This discrepancy illustrates the important point that, given the large scatter in all of our correlations, 
{\it correlation is not commutative.}  That is, while there is a general trend that Ly$\alpha$ emission strength increases for more nucleated objects,
this nucleated population is not the same as that moderately nucleated, large $I$ population for which $\Delta v_{\rm em - abs}$ peaks.

\subsection{Rest-Optical Spectroscopic Features}

\begin{figure}
\plotone{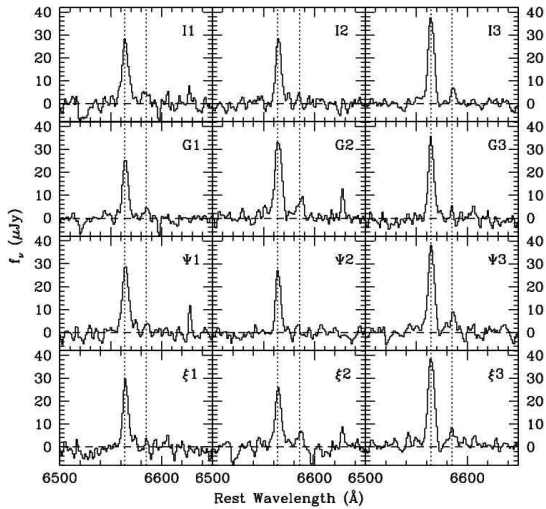}
\caption{Flux-calibrated, composite H$\alpha$ spectra are plotted for each of three bins in the morphological parameters
$I$ (size), $G$ (gini), $\Psi$ (multiplicity), and $\xi$ (color dispersion).  Spectra are numbered according to increasing values
of the morphological parameter (e.g. $G3$ is composed of more nucleated sources than $G1$).  The fiducial wavelength of H$\alpha$
and [N\,{\sc ii}] $\lambda 6583$\AA are indicated by dotted lines.  Note that the strength of H$\alpha$ 
and [N\,{\sc ii}] is not significantly different in
any of these panels.}
\label{morph_hacomp.fig}
\end{figure}

As part of an ongoing near-IR spectroscopic survey (Erb et al. 2006a) we have obtained rest-frame optical spectra in the wavelength regime
of H$\alpha$ and [N\,{\sc ii}] for 19 of the 150 galaxies in the $z \sim 2$ galaxy sample, and use these spectra to explore whether there is any
apparent relation between morphology and H$\alpha$ flux and/or 
the oxygen abundance as measured by the [N\,{\sc ii}]/H$\alpha$ ratio (Pettini \& Pagel 2004).
Given the extremely small sample of galaxies with near-IR spectra, we divide these 19 galaxies into only three bins according to our
morphological parameters and analyze the resulting stacked spectra (shown in Figure \ref{morph_hacomp.fig}) 
with a method similar to that adopted for the rest-UV spectra.

\begin{figure}
\plotone{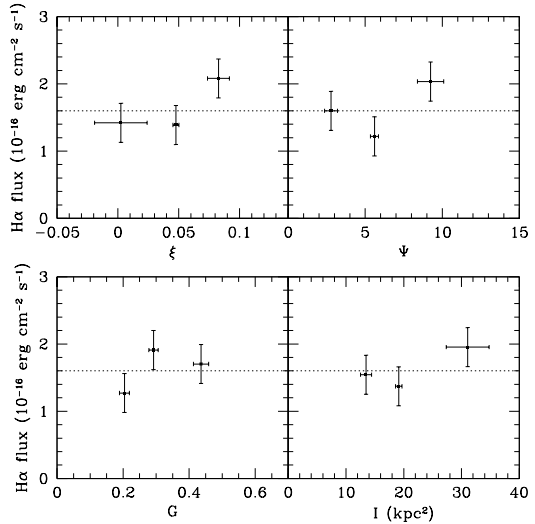}
\caption{H$\alpha$ fluxes measured for the composite stacks in each of the morphological parameters $I$ (size), 
$G$ (gini), $\Psi$ (multiplicity), and $\xi$ (color dispersion).  Horizontal errorbars represent uncertainty in the mean value
of a morphological quantity within each bin, vertical errorbars represent $1\sigma$ uncertainties on the flux based on
random-draw tests.  The dotted line in each panel represents the mean H$\alpha$ flux of the entire sample.}
\label{morph_hameas.fig}
\end{figure}

As indicated by the general similarity of all of the composites shown in Figure \ref{morph_hacomp.fig}, and plotted more precisely in
Figure \ref{morph_hameas.fig}, we find no significant variation in the strength of H$\alpha$ emission with rest-UV morphology.
[N\,{\sc ii}] is only marginally detected in many of the composite spectra, and all variations are well within the uncertainty expected based
on the noise of the composite spectra.
We conclude therefore that to within the accuracy permitted by our small (and hence not fully representative of the large
distribution of UV morphologies) spectroscopic sample the UV morphology of $z \sim 2$ galaxies is uncorrelated with rest-frame optical
spectroscopic features and the degree of metal enrichment as parametrized by the [N\,{\sc ii}]/H$\alpha$ oxygen abundance estimate.


\section{THE ASSOCIATION OF REST-UV MORPHOLOGIES WITH PHOTOMETRICALLY DERIVED PROPERTIES}

Using ground-based $U_nG{\cal R}$ and {\it Spitzer}- MIPS photometry it is possible to calculate the rest-frame, $k$-corrected
luminosities of each of our target galaxies in the UV ($L_{1600}$) and mid-IR ($L_{5-8 \mu}$)
\footnote{$L_{1600}$ typically ranges from $\sim 10^{10} L_{\odot} - 1.2 \times 10^{11} L_{\odot}$, $L_{5-8 \mu}$
from $\sim 5 \times 10^{9} L_{\odot} - 1.3 \times 10^{11} L_{\odot}$; see Reddy et al. 2006a for further details}
, in addition to estimating the total bolometric
luminosity ($L_{\rm BOL}$) and the ratio of IR/UV luminosities ($L_{\rm FIR}/L_{\rm 1600}$).
The resulting spectral energy distribution from these and additional $JK$ and {\it Spitzer}- IRAC data may then be fit with stellar population models to 
determine the best-fit stellar mass ($M_{\star}$), age, and optical extinction ($E(B-V)$) for a given galaxy.
Although the risk of confusion is greater in stellar population models based on such seeing-limited imaging, 
in almost all cases (except of those of the few most widely separated clumps) we find that the isophotes of the target galaxies
reliably trace the ACS morphology, and all components are blurred together into a single object with minimal contamination from nearby sources.
The comprehensive results of such efforts have been summarized by Reddy et al. (2006a); in the present contribution
we test the degree of association between morphology and such photometrically-derived parameters.
In this section we consider only the 62\% of the $z \sim 2$ sample of galaxies
for which {\it Spitzer}-MIPS 24$\micron$ detections directly measure the strength of rest-frame mid-IR emission (see further
discussion in Reddy et al. 2006b).  

We quantify the degree of association using the Spearman 
non-parametric rank
correlation coefficient $r_s$, values of which are assessed
in terms of their two-sided significance $t$.  This significance gauges the probability that the null hypothesis (i.e. that
there is no correlation between the two parameters) is true and that any apparent correlation is due to random chance alone.
In Table \ref{spearman.table} we calculate $t_{\sigma}$
\footnote{Since $t$ is an awkward statistic to print for extremely small values, we instead give results in terms of $t_{\sigma}$ corresponding
to the number of sigma that $t$ lies out along the wings of a gaussian probability distribution.}
for a grid of photometric versus morphological parameters (adopting the convention that positive
$t_{\sigma}$ indicates a positive correlation, and negative $t_{\sigma}$ a negative), highlighting
in bold typeface those for which the null hypothesis is less than 10\% probable (i.e. $|t| < 0.1$, $|t_{\sigma}| > 1.65$).
We note from Table \ref{spearman.table} that the null hypothesis is rejected in very few cases, indicating that in general morphology is largely 
decoupled from photometrically derived parameters.

\begin{deluxetable}{lcccc}
\tablecolumns{5}
\tablewidth{0pc}
\tabletypesize{\scriptsize}
\tablecaption{Standard deviations from the null hypothesis for independence between morphological and photometric parameters.\tablenotemark{a}}
\tablehead{
\colhead{Quantity} & \colhead{$I$} & \colhead{$G$} & \colhead{$\Psi$} &\colhead{$\xi$}}
\startdata
$L_{1600}$\tablenotemark{b} & \boldmath $+6.39$ & \boldmath $+4.57$ & $-0.32$ & \boldmath $+3.10$\\
$L_{5-8 \mu}$\tablenotemark{c} & $+0.92$ & $-0.10$ & $+0.68$ & $-0.30$\\
$L_{5-8 \mu}$\tablenotemark{d} & $+0.21$ & $+0.59$ & $+0.02$ & $+0.68$\\
$L_{\rm FIR}/L_{1600}$\tablenotemark{e} & \boldmath $-3.33$ & \boldmath $-1.85$ & $-0.22$ & $-1.22$\\
$L_{\rm BOL}$\tablenotemark{f} & $+0.71$ & $+0.90$ & $+0.03$ & $+0.73$\\
$M_{\star}$\tablenotemark{g} & $+0.98$ & $-0.05$ & $-0.99$ & $+0.07$\\
Age\tablenotemark{h} & $-0.83$ & $-0.02$ & $-0.16$ & $+1.12$\\
$E(B-V)$\tablenotemark{i} & \boldmath $-3.29$ & \boldmath $-2.35$ & $-0.68$ & \boldmath $-1.84$\\
\enddata
\label{spearman.table}
\tablenotetext{a}{`+' denotes positive correlation, `-' denotes negative.  Bold typeface indicate correlations of greater than 90\% confidence.}
\tablenotetext{b}{Rest-frame ultraviolet luminosity.}
\tablenotetext{c}{Rest-frame mid-IR luminosity (all sources).}
\tablenotetext{d}{Rest-frame mid-IR luminosity (MIPS sources only).}
\tablenotetext{e}{Ratio of far-IR to ultraviolet luminosities.}
\tablenotetext{f}{Bolometric luminosity.}
\tablenotetext{g}{Stellar mass.}
\tablenotetext{h}{Age of stellar population.}
\tablenotetext{i}{Estimated reddening based on far-UV spectral slope.}
\end{deluxetable}

The most significant correlations relate galaxy size ($I$), nucleation ($G$),
and rest-frame ultraviolet luminosity ($L_{\rm 1600}$) in
the sense that larger and more strongly nucleated galaxies tend to have brighter UV continua.
In general, these correlations are unsurprising
since $I$ effectively measures the number of UV bright pixels in a galaxy (and a galaxy with
large $I$ is hence likely to have a large total UV luminosity), and high values of $G$ are produced by relatively high concentrations of flux in a small
percentage of the total number of UV bright pixels (which occurs more commonly in galaxies with a high total UV luminosity to distribute).
However, we note that the positive correlation between $G$ and $L_{\rm 1600}$ may also (at least in part) have a more physical explanation; if indeed
nebulous low $G$ sources are dustier, then UV radiation from these galaxies would be more strongly attenuated than from less dusty high $G$ galaxies,
contributing to the observed trend.

The next most significant correlations relate $I$, $G$, 
$E(B-V)$ (the estimated reddening based on stellar population models and the far-UV spectral slope)
and the ratio of IR to UV luminosity ($L_{\rm FIR}/L_{1600}$, a proxy
for UV attenuation).  As quoted in
Table \ref{spearman.table}, $I$ and $G$ are both negatively correlated with $E(B-V)$ and $L_{\rm FIR}/L_{1600}$ (at levels of significance ranging from
about $2 - 3 \sigma$) in the sense that
 the most attenuated galaxies tend to have 
smaller UV luminous areas (low $I$) with weaker nucleation in the flux distribution (low $G$).
Although this correlation contains a great deal of scatter and does not hold precisely for every galaxy 
(i.e. not all ``small and nebulous'' galaxies are highly extinguished), it nonetheless appears
that {\it statistically} the most extinguished galaxies tend to be smaller and more nebulous than the general population.
As previously hypothesized on the basis of Ly$\alpha$ emission strength, it may therefore be the case that galaxy morphology is dictated 
in part by the presence
and location of dust which partially extinguishes the UV continuum radiation and causes galaxies to appear slightly smaller (to a fixed limiting surface
brightness) and more nebulous.

The color dispersion $\xi$ appears to follow the trends found for $I$ and $G$ with UV luminosity and extinction (although at lower significance),
possibly due to its known positive correlation with $G$.  The multiplicity parameter $\Psi$  however (also loosely correlated with $G$) exhibits no
statistically significant associations with any of the photometrically derived quantities presented in Table \ref{spearman.table}.
In addition, we note that contrary to possible expectations (e.g that galaxies with higher stellar mass might, analagously to local
elliptical galaxies, appear morphologically more evolved and distinct from those with low stellar mass)
there is no evidence for a relationship between stellar mass and UV morphology, 
implying that through a combination of physical processes galaxies ranging over two decades in stellar
mass from $M_{\star} \sim 10^9 - 10^{11} M_{\odot}$ somehow appear visually indistinguishable.
However, while no significant {\it overall}
correlation was found by the Spearman test, almost all (20/22) of the 
most strongly nucleated sources ($G > 0.4$) have masses greater than
$10^{10} M_{\odot}$ in contrast to those with $G < 0.4$ which span a full two decades in stellar mass.

\begin{deluxetable}{lcccc}
\tablecolumns{5}
\tablewidth{0pc}
\tabletypesize{\scriptsize}
\tablecaption{Standard deviations from the null hypothesis for independence of morphological parameters and star formation.\tablenotemark{a}}
\tablehead{
\colhead{Quantity} & \colhead{$I$} & \colhead{$G$} & \colhead{$\Psi$} &\colhead{$\xi$}}
\startdata
SFR$_{\rm SED}$\tablenotemark{b} & $-0.32$ & $+0.59$ & $-0.15$ & $+0.30$\\
SFR$_{\rm UV}$\tablenotemark{c} & \boldmath $+8.04$ & \boldmath $+4.77$ & $+0.46$ & \boldmath $+3.15$\\
SFR$_{\rm IR}$\tablenotemark{d} & $+0.21$ & $+0.59$ & $+0.02$ & $+0.47$\\
SFR$_{\rm UV+IR}$\tablenotemark{e} & $+0.92$ & $+1.05$ & $+0.03$ & $+0.87$\\
SFR$_{\rm UV+IR}$\tablenotemark{f} & \boldmath $+5.67$ & \boldmath $+2.17$ & $+0.80$ & \boldmath $+1.76$\\
$\phi$\tablenotemark{g} & $+1.24$ & $+1.62$ & $+0.88$ & $+1.01$\\
$\phi$\tablenotemark{h} & $-1.12$ & $-0.43$ & $-0.04$ & $-0.70$\\
\enddata
\label{spearmanSFR.table}
\tablenotetext{a}{'+' denotes positive correlation, '-' denotes negative.}
\tablenotetext{b}{Star formation rate estimated from SED fitting.}
\tablenotetext{c}{Star formation rate estimated from UV photometry.}
\tablenotetext{d}{Star formation rate estimated from IR photometry.}
\tablenotetext{e}{Star formation rate estimated from UV+IR photometry, MIPS detected sources only.}
\tablenotetext{f}{Star formation rate estimated from UV+IR photometry, MIPS undetected sources only.}
\tablenotetext{g}{Specific star formation rate, $\phi =$ SFR$_{\rm UV+IR}/M_{\star}$.  MIPS detected sources only.}
\tablenotetext{h}{Specific star formation rate, $\phi =$ SFR$_{\rm UV+IR}/M_{\star}$.  MIPS undetected sources only.}
\end{deluxetable}

In Table \ref{spearmanSFR.table} we compare rest-UV morphologies to star formation rate (SFR) estimates based on
SED fitting, UV and IR continuum emission.  Although these estimates tend to be loosely correlated
with each other, there is considerable scatter in individual estimates of the SFR for a particular
source using different methods.  
The least reliable of our estimates are the SFR estimated from
normalization of stellar population models (Shapley et al. 2005; Erb et al. 2006a; Reddy et al. 2006a), for which we find no significant correlations with
morphological parameters (due either to the intrinsic lack of such a correlation, or to the uncertainty inherent in the SFR estimate given by this
method).  The next most reliable estimate is that based on UV continuum flux, for which we calculate the uncorrected star formation
rate based on $U_nG{\cal R}$ photometry and the conversion factor given by Kennicutt (1998).  
As shown in Table \ref{spearmanSFR.table},
the uncorrected UV SFR is strongly correlated with all morphological parameters save the multiplicity $\Psi$.  However, this is not surprising given
the strong association between total UV luminosity and morphology shown in Table \ref{spearman.table}.
While it is likely that large (high $I$),
nucleated (high $G$) objects have a higher than average 
rate of {\it uncorrected} star formation it is not necessarily true that the same holds for the total star formation rate.
Indeed, in the majority of cases {\it obscured} star formation comprises the bulk of the total.
Rather than assuming an average extinction factor to estimate
this from the UV luminosity, we estimate the obscured SFR from the IR luminosity $L_{\rm IR}$ calculated by Reddy et al. (2006b), adopting the
$L_{\rm IR} -$ SFR
conversion factor given by Kennicutt (1998).  As shown in Table \ref{spearmanSFR.table}, obscured star formation has far less compelling
associations with UV morphology- none are found to have statistical significance.

The most accurate estimate of the total star formation rate may be found by summing the rates determined from UV and IR
luminosities.  We break this comparison into two cases- galaxies for which the IR-derived SFR has been directly measured using {\it Spitzer}-MIPS detections at
24$\micron$, and galaxies which are undetected at 24$\micron$ and an estimate of the IR-derived SFR has been computed based upon the average bolometric luminosity
of MIPS-undetected sources\footnote{$L_{\rm IR}$ effectively scales with $L_{\rm BOL}$.}
($L_{\rm BOL} = 6 \times 10^{10} L_{\odot}$, Reddy et al. 2006b).
For the first of these cases, no significant correlations are found with either the total or specific (stellar mass-normalized) star formation rate.
We note however that the specific star formation rate is on the threshhold of statistical significance for a positive correlation with $G$.
If the eleven galaxies with stellar populations younger than 40 Myr (i.e. for which stellar population fits are the most uncertain)
are neglected in our analysis, this correlation becomes statistically significant, deviating from the null hypothesis by approximately $2.1 \sigma$.
While therefore there is mild evidence for a positive correlation between
UV nucleation and net star formation for galaxies directly detected at 24$\micron$, this association is not at present statistically compelling.
In contrast, for the $24\micron$-undetected sources all of the associations between morphology and uncorrected UV SFR are recovered for the total estimated SFR
(although this trend is smeared out by the division by mass for the specific SFR), largely
because the obscured SFR extrapolated from total bolometric luminosity
contributes only marginally to the total in these cases.  



\section{A COMPARISON OF MORPHOLOGIES WITH OTHER GALAXY SAMPLES}

\subsection{AGN/QSO}

\begin{figure}
\plotone{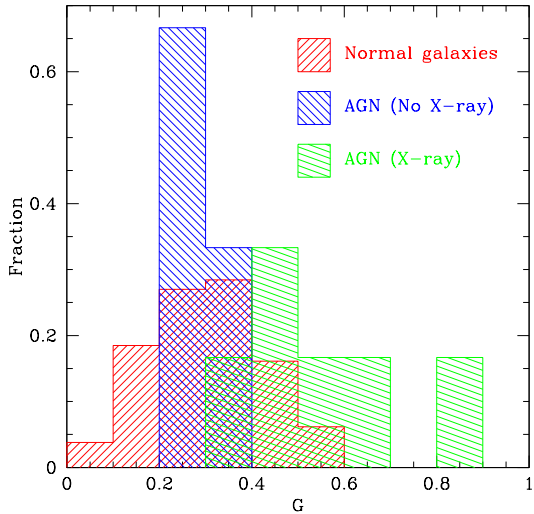}
\caption{Histrogram showing the relative UV nucleation for the $z = 1.8-3.4$ galaxy sample (red histogram) versus
X-ray detected (green histogram) and X-ray undetected AGN/QSO (blue histogram).  The nucleation of the X-ray undetected sample
is broadly consistent with the median nucleation of the overall $z = 1.8 -3.4$ population, while X-ray detected galaxies are more strongly
nucleated.}
\label{gini.agn.fig}
\end{figure}

As given in the table of AGN/QSO in GOODS-N presented by Reddy et al. (2006a, their Table 3),
there are a total of nine spectroscopically confirmed AGN/QSO in our sample in the redshift range $z = 1.8 - 3.4$,
six of which are directly detected in X-ray emission (although one very weakly) and three of which have no
X-ray counterpart to a depth of 2 Ms (based on the catalog of Alexander et al. 2003) but are confirmed AGN based upon 
high-ionization optical emission lines and/or
power-law mid-IR SEDs.
As noted by Reddy et al. (2006a), 
the morphologies of the X-ray undetected sample are more disturbed than their
directly detected counterparts.  As we show in Figure \ref{gini.agn.fig}, all three
X-ray undetected sources have $G<0.4$, while the the five detected sources all have strongly nucleated values $G > 0.38$ 
(four with $G > 0.47$) and the two QSO in the sample
(BMZ 1083 and MD 39) have the highest values of any object considered in this contribution
at $G = 0.63$ and $0.88$ respectively (close to the stellarity limit
of $G \sim 0.90$).
This suggests a possible correlation between X-ray luminosity and UV nucleation,
in the sense that AGN which produce obvious X-ray radiation also contribute sufficiently to the total UV output of their host galaxies
that the centralized AGN radiation visibly affects the apparent nucleation of the UV light profile.
Given the extremely small size of our sample, it is difficult to asses the global applicability of this correlation although we note for comparison
that in a larger sample of 31 AGN in the redshift range $z \sim 2 - 4$ Akiyama (2005; their Fig. 3) 
found that AGN with the most centrally concentrated light distributions tended to be X-ray bright ($L_{\rm 2 - 10 keV} > 10^{44}$ erg s$^{-1}$)
broad-line sources, while the population of X-ray bright narrow-line sources and X-ray faint sources tended to be slightly less centrally concentrated.
Likewise, recent studies of AGN at lower redshifts ($z \lesssim 1.3$) by Pierce et al. (2006) and Grogin et al. (2005) also
found the rest-frame optical morphologies of X-ray luminous AGN to be more compact than those of IR-selected AGN or ordinary field galaxies.

Not only are the X-ray undetected AGN more nebulous than their X-ray bright counterparts,
we find that they also appear to have a greater number of components to their UV light,
exhibiting a range of multiplicities $\Psi = 4.6 - 11.8$ as compared to the $\Psi < 1.8$ found for all directly detected X-ray sources.
These irregular morphologies, coupled with a power-law SED longward of 3\,$\micron$ (indicating the presence of warm dust)
yet a UV SED well-fit by a simple stellar population suggest (Reddy et al. 2006b) that these sources may be obscured AGN whose UV emission is dominated
by spatially extended star formation rather than a central active nucleus.

\subsection{IR-selected $BzK$ Galaxies}

\begin{figure*}
\plotone{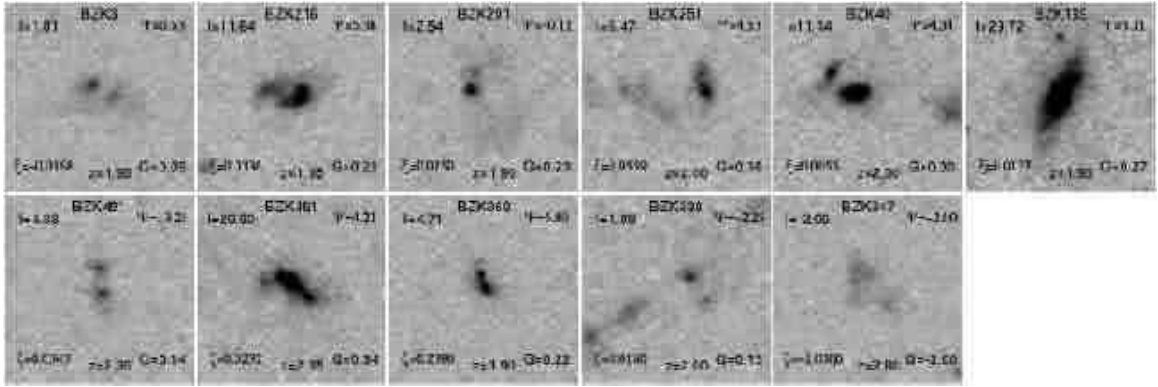}
\caption{As Figure \ref{mosaicA.fig}, but for IR-selected $BzK$ sources brighter than $K = 21$ which are not selected by the rest-UV color selection
criteria.  Redshifts given are photometric.}
\label{mosaicbzk.fig}
\end{figure*}

One well-studied near-IR color selected galaxy sample is the $BzK$- selected catalog (Daddi et al. 2004),
for which many sources (fifty two sources brighter than $K = 21$ in the redshift range $z = 1.8 - 2.6$) 
simultaneously satisfy both the $BzK$ selection criteria and our optical $U_nG{\cal R}$ color selection criteria.
In Figure \ref{mosaicbzk.fig} we show the morphologies of those GOODS-N $BzK$ galaxies which are comparably bright in $K$ band 
and and have well-determined photometric redshifts,
yet fail to meet the optical selection criteria.
Qualitatively, we note that the $BzK$ galaxies not in the optically-selected sample appear morphologically very similar to those in Figure \ref{mosaicA.fig},
as might be expected given the large general overlap between the two samples and suggesting that these few galaxies
may fall just outside of the optical selection criteria.

\begin{figure*}
\plotone{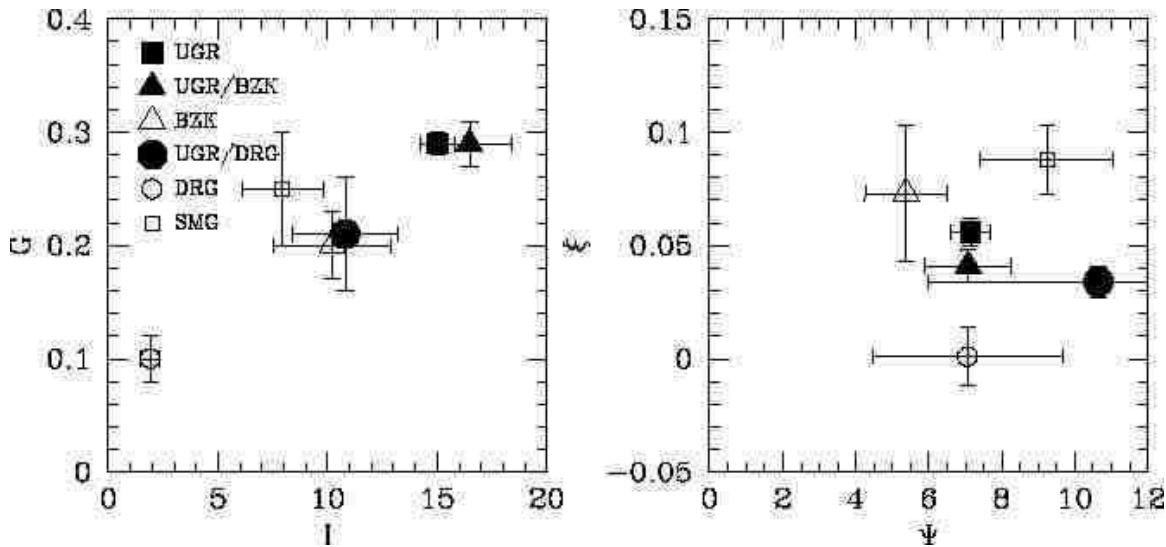}
\caption{Comparative average size ($I$), nucleation ($G$), multiplicity ($\Psi$), and color dispersion ($\xi$)
for optically selected $U_nG{\cal R}$, $BzK$, DRG, and SMG galaxy samples.  
Filled squares represent all $z \sim 2$ $U_nG{\cal R}$-selected galaxies,
filled triangles and circles the respective subsets which also fulfill $BzK$ and DRG criteria, open triangles and circles galaxies which fulfill the 
$BzK$ and DRG criteria (respectively) which do not meet the $U_nG{\cal R}$ selection criteria,
and open squares all galaxies which fulfill the SMG critera.  All galaxies are limited to the redshift range $z = 1.8 - 2.6$ (using photometric redshifts
where spectroscopic are unavailable), and the $BzK$ and DRG galaxy samples are limited those with no x-ray counterparts to 2 Ms and $K$ magnitudes
$K < 21$.  Errorbars indicate the 1$\sigma$ uncertainty in the mean values for
each population.}
\label{bzkdrg.fig}
\end{figure*}

We quantify these morphological differences in Figure \ref{bzkdrg.fig}, using photometric redshift estimates to scale the surface
brightness selection algorithm.  As expected based on Figure \ref{mosaicbzk.fig}, the average morphology
of $BzK$ sources which are also $U_nG{\cal R}$ sources (filled triangles) is identical to that of the overall $U_nG{\cal R}$ sample (filled squares)
to within the uncertainties in the mean.  The sample of eleven $BzK$ sources which do {\it not} meet the $U_nG{\cal R}$ criteria (empty triangles)
has slightly fewer high-surface brightness pixels (lower $I$) and is slightly less nucleated (lower $G$)
than the other two samples, but at a confidence level of only $1 - 2\sigma$.
The apparent multiplicity $\Psi$ and color dispersion $\xi$ remain approximately constant among all of these samples
to within the uncertainty.

\subsection{IR-selected Distant Red Galaxies}

\begin{figure*}
\plotone{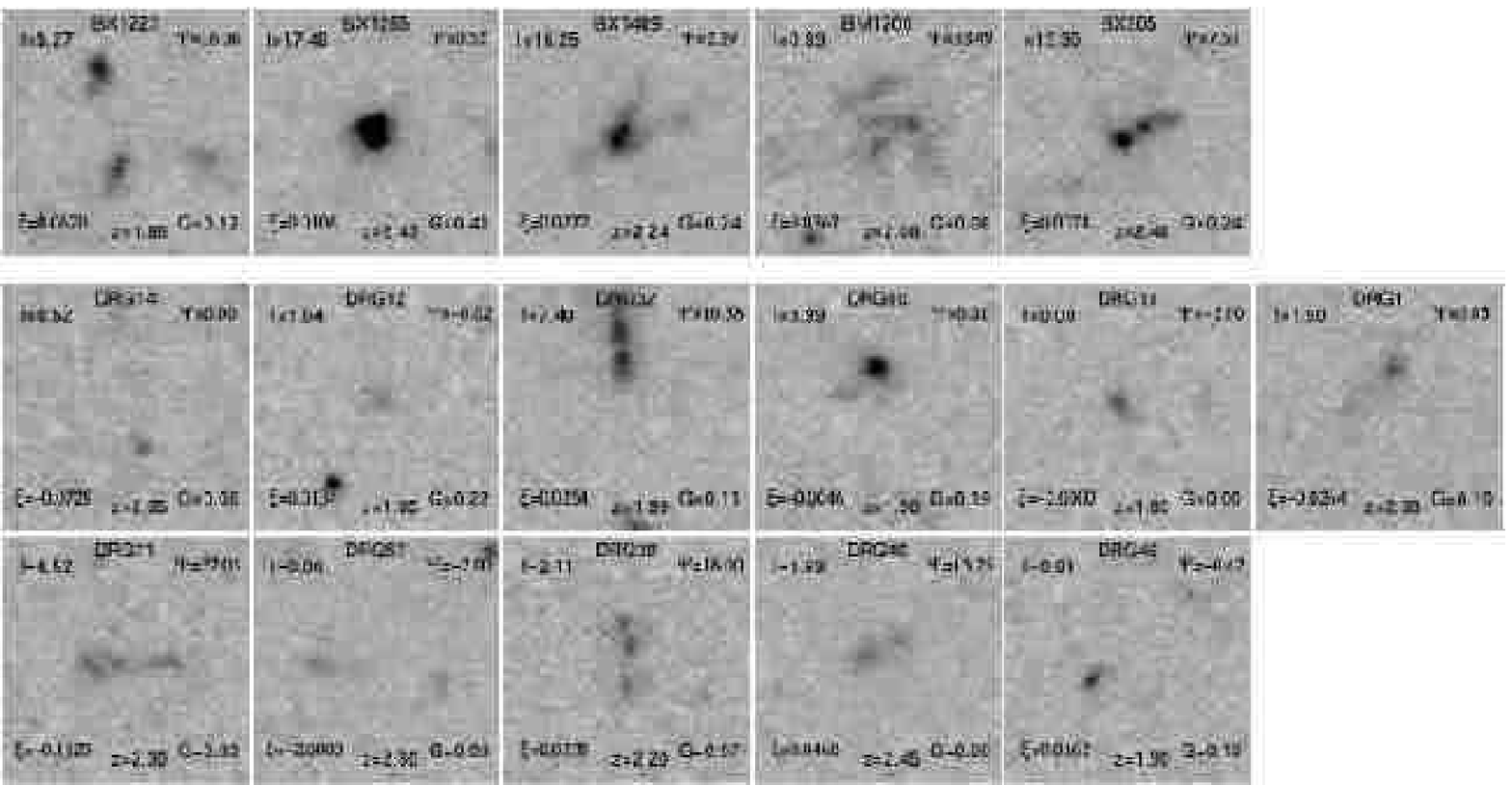}
\caption{As Figure \ref{mosaicA.fig}, but for IR-selected DRG sources brighter than $K = 21$.  Top row: DRG which are also selected by rest-UV
color selection criteria, Bottom rows: DRG which are not selected by rest-UV color selection criteria.  Redshifts given for galaxies not selected by
the rest-UV criteria are photometric.}
\label{mosaicdrg.fig}
\end{figure*}

Distant red galaxies (DRG's, Franx et al. 2003) constitute another major IR-selected sample of high redshift galaxies for which there are five sources 
brighter than $K = 21$ in the redshift range $z = 1.8 - 2.6$ which meet the $U_nG{\cal R}$ selection criteria and have
secure spectroscopic redshifts, and eleven sources which do not but for which we measure reliable photometric redshifts. 
As illustrated by Figure \ref{mosaicdrg.fig}, the DRG population is fainter and more nebulous in UV emission than any of the other
galaxy samples considered, consistent with what may be expected for particularly dusty IR-bright galaxies given the
apparent correlation between UV nucleation and dust extinction.

As for the $BzK$ sample, the overlapping sample of $U_nG{\cal R}$-selected DRGs has morphological coefficients similar to
the bulk of the $U_nG{\cal R}$ population to within $1 - 2\sigma$ (although it is on average
slightly slightly smaller and more nebulous, see Fig. \ref{bzkdrg.fig}).
It is those DRG {\it not} selected by $U_nG{\cal R}$ criteria however which particularly stand out-
these galaxies (Fig. \ref{bzkdrg.fig}, open circles) have
much smaller $I$ and $G$ to a high level of confidence and in some cases (e.g. DRG 14) are barely detected in our {\it HST}-ACS imaging data.
Largely, this difference is a reflection of the much fainter UV luminosity of the DRG sample-
typical DRG which do not meet the optical selection criteria have
UV luminosities $L_{1600} < 10^{10} L_{\odot}$ (and commonly $L_{1600} < 10^{9} L_{\odot}$), 
as compared to the median luminosity for the optically selected sample
$L_{1600} \sim 5 \times 10^{10} L_{\odot}$.
Similarly to the $BzK$ sample, the apparent multiplicity $\Psi$ for DRGs is again consistent with the $U_nG{\cal R}$ sample, although
the color dispersion $\xi$ is roughly $2\sigma$ lower
and is consistent with zero (the value expected for a pure measurement of the background sky).

\subsection{Submillimeter Galaxies}

\begin{figure*}
\plotone{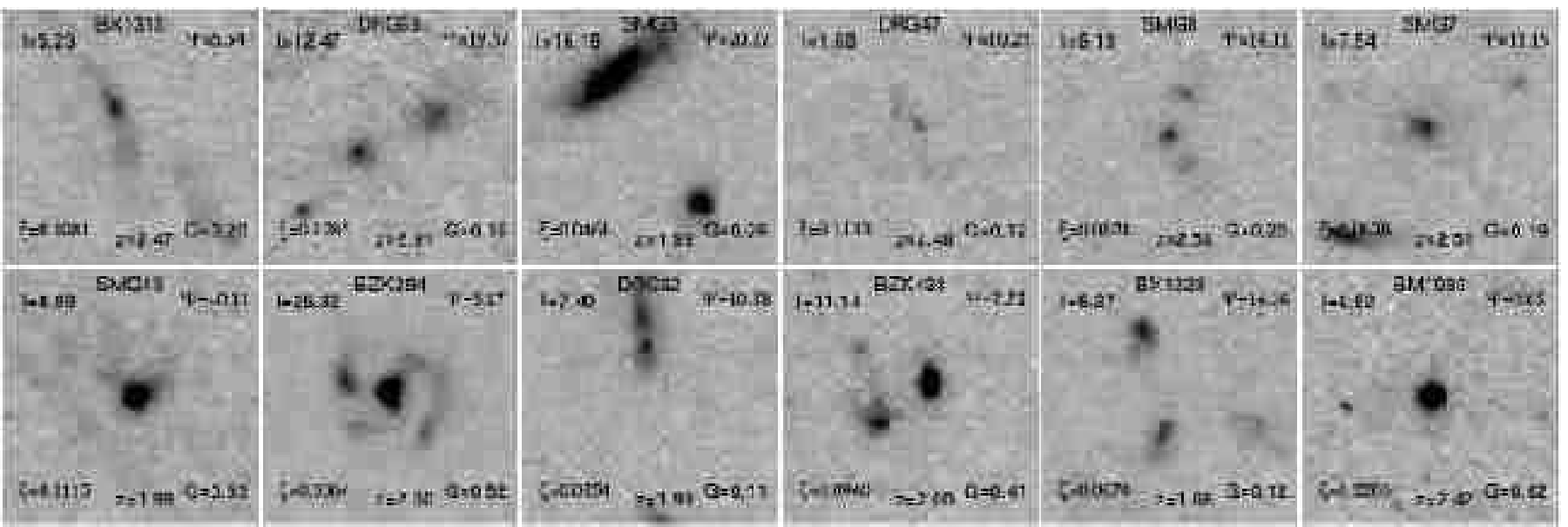}
\caption{As Figure \ref{mosaicA.fig}, but for all SMGs on the HST-ACS image frame of the GOODS-N field in the redshift range $z = 1.8 - 2.6$
(plus one source at $z = 2.91$).  Positions and spectroscopic redshifts are drawn from the catalog of Chapman et al. (2005).
Galaxies which also fulfill $U_nG{\cal R}$, $BzK$, or DRG selection criteria are labelled accordingly.
Note that the lower right object in this figure (BM 1083) is a known QSO.}
\label{mosaicsmg.fig}
\end{figure*}

The submillimeter galaxy (SMG) population offers the opportunity to compare the rest-UV morphologies of optically-selected galaxies
with those in a similar redshift range ($z \sim 2 - 3$) which are selected on the basis of submillimeter flux.
Using the catalog of SMG coordinates and spectroscopic redshifts compiled by Chapman et al. (2005), we apply our morphological analysis
to a sample of these galaxies in the GOODS-N field.  The wide range of morphological types covered by the SMG selection criteria
are shown in Figure \ref{mosaicsmg.fig} and range from single, nucleated sources to extremely faint and nebulous, and include
one strong high-redshift spiral galaxy candidate (BZK 294- see Dawson et al. 2003).
On average however, the SMG population has a morphology similar to that of the $U_nG{\cal R}$-selected sample (Fig. \ref{bzkdrg.fig})
with values of $\Psi$ and $\xi$ consistent to within the uncertainty and apparent sizes $I$ and nucleations $G$ only slightly smaller and
more nebulous than the $U_nG{\cal R}$-selected sample.  These results are consistent (since our size parameter $I$ is
closely related to surface brightness) with the findings of
Chapman et al. (2003) and Conselice et al. (2003b), whose {\it HST}-STIS imaging led them to conclude that
the SMG sample had lower surface brightness than typical $z \sim 3$ LBGs, yet was slightly larger {\it for the reduced surface brightness}.

Additionally, Conselice et al. (2003b) also find that the asymmetry index $A$ of their sample of eleven SMGs is slightly greater on average
than that of optically-selected $z \sim 3$ LBGs, which these authors conclude implies a greater major merger fraction for the submillimeter-bright sample.
However, we find no statistically significant change in our multipliticy parameter $\Psi$ (which most closely measures the morphological irregularity
of a source).
If the morphologies of the SMG sample are governed by the same physical processes as the optically-selected galaxies (as seems likely
given the two samples appear to fall on the same trend of dustiness versus bolometric luminosity
for star-forming galaxies at $z \sim 2$; Reddy et al. 2006b) then
when considered non-parametrically we conclude that the SMG sample may be no more likely
to be dominated by major mergers than other $z \sim 2 - 3$ galaxy samples.

\section{DISCUSSION}

Through a comprehensive statistical analysis, we have found two key trends with morphology.  First, the degree of nucleation/nebulosity
is related to the UV/optical extinction of the source (as parametrized through both the value of $E(B-V)$ calculated from stellar population
models and the ratio of IR to UV luminosity), suggesting that more nebulous sources may appear as such since their greater quantities
of dust obscure a great quantity of the UV light.  Second, physically larger, more UV luminous sources 
have rest-UV emission and absorption lines seperated by a greater velocity, indicating that they may
drive more energetic outflows than their lower luminosity counterparts.
However, we note that these trends are typically on the order of $3 - 4\sigma$ significance,
and are far from being unambiguously
convincing morphological differences.  Indeed, given the relatively large sample of galaxies and volume of spectrophotometric data
compiled to date, the overall {\it lack} of correlation between morphology and fundamental galactic properties
such as stellar mass, SFR, and outflows is more remarkable.  To some degree this may be due to the unknown distribution of viewing angles, but whether
as a result of this complication or a more fundamental process the rest-UV morphologies of high redshift galaxies
generally do not appear to contain a great deal of separable information.  If morphologies truly mean so little, this may help explain
the lack of correlation seen between elongated morphologies and major axis velocity shear by Erb et al. (2004), who found
that morphologically elongated galaxies are no more likely than compact galaxies to 
exhibit kinematic signatures of rotation.

It is interesting to note the implications of these findings for the major-merger hypothesis which posits that,
similarly to luminous and ultraluminous infrared galaxies (LIRGs and ULIRGs) in the local universe, 
particularly irregular morphologies correspond to major galaxy mergers (e.g. Conselice et al. 2003, Chapman et al. 2003, LPM04, Lotz et al. 2006).
In particular, Conselice et al. (2003a) adopt the asymmetry selection criterion $A > 0.35$ to identify major mergers, 
and interpret the increase in galaxy
irregularity with redshift out to $z \sim 3$ to indicate a higher merger fraction in the early universe and evidence for hierarchical formation.
If the most irregular galaxies truly represent such major events however, we might
reasonably expect these galaxies to differ significantly from their non-merging counterparts in some additional way, perhaps by displaying
tidally enhanced rates of star formation or broadened interstellar absorption lines due to the juxtaposition of the ISM of two or more
galaxies with relative velocity differences of order a few hundred km s$^{-1}$.
However, we find no evidence for such a distinction between any of a variety of morphological samples, suggesting that either
major mergers at $z \sim 2 - 3$ are largely indistinguishable from the non-merging sample, 
that {\it all} $z \sim 2 - 3$ galaxies are experiencing major merger induced starbursts, or simply that
irregular morphology (at least to within $\sim$ 13 kpc at rest-UV wavelengths) is not a reliable indicator of a major merger.
While it is possible that major mergers may be distinguishable on the basis of multiple components beyond our 13 kpc detection
threshhold, it seems unlikely that such systems would produce changes as profound as may be found on closer passes.

Supposing then that irregular, multi-component morphologies do {\it not} represent bursts of star formation in the orbiting
galaxies and satellite galaxies of a major merger, another possibility is that these systems may instead 
represent patchy star formation occuring from the collapse of local instabilities within galactic molecular clouds.
Combined with the complicating effects of dust and viewing angle, such an explanation may fit the local LBG-analog
VV 114, based on the UV and near-IR imaging of Goldader et al. (2002).
It is also possible that, in some small number of cases, multi-component morphologies may reflect objects which are nearby
in projection, but unrelated due to large differences in redshift.  While individual components of multi-component
objects have not been spectroscopically confirmed to be associated however, the angular distribution of sources suggests that such
projection effects should contribute minimally to the total population.

Finally, we note that while the rest-UV and rest-optical morphologies of $z \sim 2 - 3$ galaxies are similar in the majority of cases
(as shown by Dickinson 2000)
and appear to contain little separable information, there is a small sub-sample of these
galaxies for which the optical morphologies appear considerably more regular and ``evolved'' than their UV counterparts (Toft et al. 2005).
It may be the case that this small population of galaxies has rest-optical morphologies which are easier to interpret
than the majority of the galaxy sample, and may have useful correlations between rest-optical morphology and spectrophotometric
or kinematic indices.  In particular, such galaxies are intriguing possible targets for future kinematic study
with the aid of AO-assisted integral-field spectroscopy (e.g. Forster-Schreiber et al. 2006; Law et al. 2006),
which could demonstrate whether these galaxies are any more likely than the rest of the galaxy sample to 
contain tractable meaning in their luminous spatial structures.

\acknowledgements
DRL and CCS have been supported by grant
AST03-07263 from the US National Science Foundation, and archival grant HST-AR 10311 from the Space Telescope Science Institute.
The authors thank Rupali Chandar and Christy Tremonti for helpful communications.

\end{document}